\documentclass[useAMS,usenatbib]{mn2e}
\usepackage{amsmath,amssymb}
\usepackage{epsfig}
\usepackage{colortbl}
\newcommand{\Dnu}{\Delta \nu}
\newcommand{\Dnusun}{\Delta \nu_\odot}
\newcommand{\nmax}{\nu_\mathrm{max}}
\newcommand{\nmaxsun}{\nu_\mathrm{max,\odot}}

\newcommand{\abolsun}{A_\mathrm{bol,\odot}}
\newcommand{\muhz}{\mu \mbox{Hz}}
\newcommand{\teff}{T_\mathrm{eff}}
\newcommand{\teffsun}{T_\mathrm{eff,\odot}}
\newcommand{\tauosc}{\tau_\mathrm{osc}}
\newcommand{\tauoscsun}{\tau_\mathrm{osc,\odot}}

\title[A Bayesian approach to amplitudes of solar-like oscillations]{A Bayesian approach to scaling relations for amplitudes of solar-like oscillations in \textit{Kepler} stars}
\author[E. Corsaro et al.]{E. Corsaro,$^{1,2,3}$\thanks{E-mail: eco@oact.inaf.it;}  H.-E.~Fr\"{o}hlich,$^{4}$ A. Bonanno,$^2$ D. Huber,$^5$ T.~R.~Bedding,$^{6,7}$
\newauthor O. Benomar,$^{6,7}$ J. De Ridder$^3$, and D. Stello$^{6,7}$\\
$^{1}$Department of Physics and Astronomy, Astrophysics Section, University of Catania, Via S. Sofia 78, I-95123 Catania, Italy\\
$^{2}$I.N.A.F. - Astrophysical Observatory of Catania, Via S. Sofia 78, I-95123 Catania, Italy\\
$^{3}$Instituut voor Sterrenkunde, K.\,U. Leuven, Celestijnenlaan 200D, 3001 Leuven, Belgium\\
$^{4}$Leibniz Institute for Astrophysics Potsdam (AIP), An der Sternwarte 16, 14482 Potsdam, Germany\\
$^{5}$NASA-Ames Research Center, Moffett Field, CA 94035-0001, USA\\
$^{6}$Sydney Institute for Astronomy (SIfA), School of Physics, University of Sydney, NSW 2006, Australia\\
$^{7}$Stellar Astrophysics Centre, Department of Physics and Astronomy, Aarhus University, Ny Munkegade 120, DK-8000 Aarhus C, Denmark}
\begin{document}

\date{Accepted . Received ; }

\pagerange{\pageref{firstpage}--\pageref{lastpage}} \pubyear{2012}

\maketitle

\label{firstpage}

\begin{abstract}
We investigate different amplitude scaling relations adopted for the asteroseismology of stars that show solar-like oscillations. Amplitudes are among the most challenging asteroseismic quantities to handle because of the large uncertainties that arise in measuring the background level in the star's power spectrum. We present results computed by means of a Bayesian inference on a sample of 1640 stars observed with \textit{Kepler}, spanning from main sequence to red giant stars, for 12 models used for amplitude predictions and exploiting recently well-calibrated effective temperatures from SDSS photometry. We test the candidate amplitude scaling relations by means of a Bayesian model comparison. We find the model having a separate dependence upon the mass of the stars to be largely the most favored one. The differences among models and the differences seen in their free parameters from early to late phases of stellar evolution are also highlighted.
\end{abstract}


\section{Introduction}
\label{sec:intro}
The study of solar-like oscillations \citep[e.\,g. see][for summaries and reviews]{BK03,BK08,BeddingWS} has experienced an enormous growth in the last five years thanks to the launch of the photometric space-based missions \textit{CoRoT} \citep{Baglin06,Michel08} and \textit{Kepler} \citep{Borucki10,Koch10}. The latter in particular, is providing a very large amount of high quality light curves, with a very high duty cycle \citep[see][for a general introduction to the \textit{Kepler} asteroseismic program, for the presentation of the \textit{Kepler} Input Catalog, and for a description of the preparation of \textit{Kepler} light curves, respectively]{Gilliland10b,Brown11,Garcia11}. These will become longer with the upcoming era of the \textit{Kepler} extended mission \citep{Still12}. 

These asteroseismic studies have recently led to the birth of the ensemble asteroseismology \citep{Chaplin11sc}, which is showing great potential for a thorough understanding of stellar evolution theory. The success of ensemble asteroseismology relies mainly on adopting scaling relations: generally simple empirical laws that allow for the derivation of fundamental stellar properties for stars different from the Sun by scaling their asteroseismic quantities from the solar values.

Among the most challenging asteroseismic quantities to measure and model one can certainly mention the oscillation amplitude. This is due to both the difficulty in estimating the background level in the power spectrum and the rather complicated physics involved in the driving and damping mechanisms of the modes \citep[e.\,g. see][]{K05,K08}. Different scaling relations aimed at predicting amplitudes by scaling from the Sun's values have been derived and discussed by several authors, both theoretically \citep{KB95,Houdek99,Houdek02,Houdek06,Samadi07,Belkacem11a,KB11,Samadi12} and observationally \citep{Stello11,Huber11,Mosser12}. A variety of amplitude scaling relations has been used extensively in literature in both ensemble studies \citep[e.\,g. see][]{Stello10,Chaplin11,Verner11,Huber11,Mosser12,Belkacem12b} and detailed analyses of single stars from main sequence to the subgiant phase of evolution \citep[e.\,g.][]{Bonanno08,Samadi10,Huber11a,Mathur11,Campante11,Corsaro12a}.

The underlying physical meaning of these various amplitude scaling relations is still not properly understood \citep[e.\,g. see the discussions by][]{Samadi07,Verner11,Huber11,Samadi12,Belkacem12b}. Testing them with observational data is vital for assessing the competing relations and for improving our understanding of stellar oscillations, i.\,e. the driving and damping mechanisms that produce the observed amplitudes \cite[see also][]{Chaplin11}. In this context, Bayesian methods can be of great use \citep[see e.\,g.][]{Trotta08} because they allow us to measure physical quantities of interest in a rigorous manner. Moreover, Bayesian statistics provides an efficient solution to the problem of model comparison, which is the most important feature of the Bayesian approach \citep[see also][]{Benomar09,Handberg11,Gruber12}.

In the present paper we analyze amplitude measurements of a sample of 1640 stars observed with \textit{Kepler}, together with temperature estimates derived from SDSS photometry, which we introduce in Section~\ref{sec:data}. In Section~\ref{sec:amp} we discuss the different scaling relations for predicting the oscillation amplitude per radial mode. The results obtained from a Bayesian parameter estimation for the different scaling relations are shown in Section~\ref{sec:inference} and the model comparison is presented in Section~\ref{sec:evidence}. Lastly, discussion and conclusions about the results of our analysis are drawn in Section~\ref{sec:amp_discussion} and \ref{sec:amp_conclusion}, respectively.

\section[]{Observations and Data}
\label{sec:data}
We use amplitude measurements and their uncertainties, obtained by \cite{Huber11} for a sample of 1673 stars spanning from main sequence (MS) to red giant stars (RGs) observed with \textit{Kepler} in short cadence (SC; mostly MS stars but also some subgiants and low luminosity RGs) and long cadence (LC; all RGs) modes \citep[][respectively]{Gilliland10a,Jenkins10}. Most of the 542 stars observed in SC have photometry for one month, while the 1131 stars observed in LC have light curves spanning from \textit{Kepler}'s observing quarters 0 to 6. All amplitudes were derived according to the method described by \cite{K05,K08}, which provides amplitudes per radial mode \citep[see][for more details]{Huber09}. Values of the frequency of maximum power, $\nmax$, the large frequency separation, $\Dnu$, and their uncertainties for all the stars were also taken from \cite{Huber11}, who used the SYD pipeline \citep{Huber09}.


It is important to have accurate temperature estimates for the stars of our sample. Unfortunately, those provided by the KIC \citep{Brown11} are known to suffer from significant systematic effects (see \citealt{P12} for a detailed discussion of the problem). We used revised $\teff$ derived by \cite{P12} for a total of 161977 KIC stars from Sloan Digital Sky Survey \textit{griz} filters, which were corrected using temperature estimates from infrared flux method (IRFM) $(J-K_s)$ color index for hot stars \cite[e.\,g. see][]{Casagrande10}. The revised effective temperatures are available in the online catalog \citep{Pcat12}. 

\begin{figure}
\begin{center}
\includegraphics[scale=0.71]{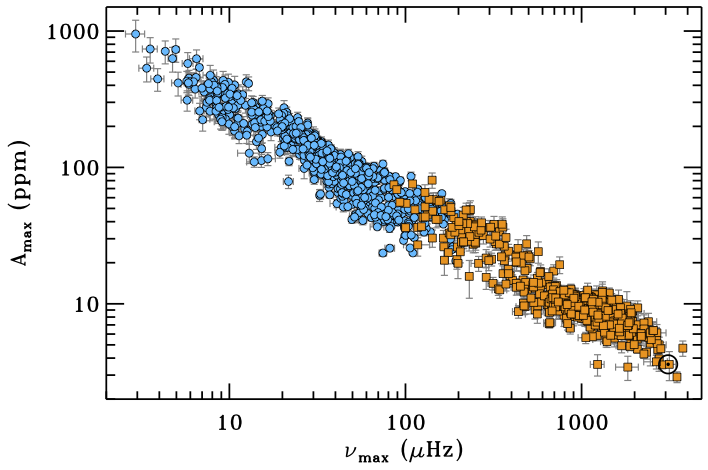}
\includegraphics[scale=0.71]{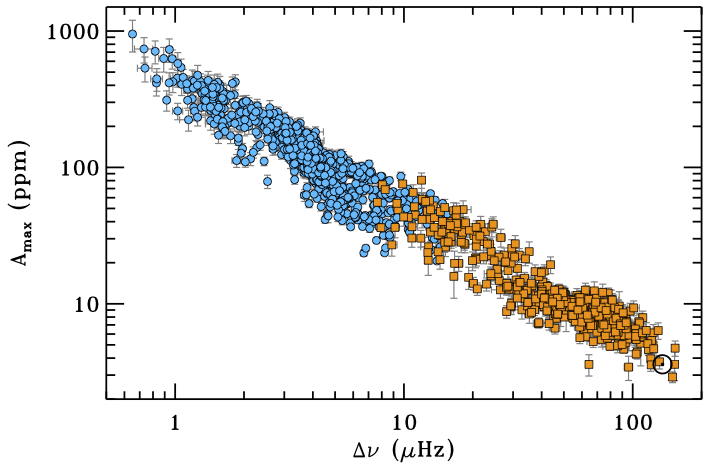} 
\includegraphics[scale=0.71]{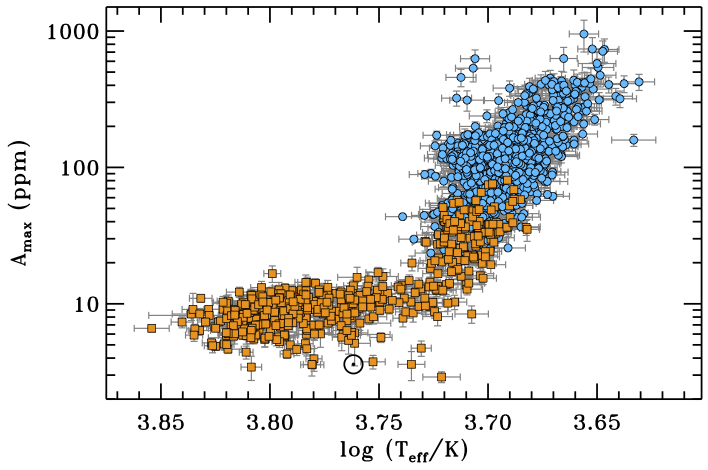}  
\caption{Oscillation amplitudes for 1640 stars observed with \textit{Kepler} in SC (orange squares) and LC (blue circles) modes and plotted against the frequency of maximum power $\nmax$ (top) and the large frequency separation $\Dnu$ (middle) of the stars in a log-log scale. Amplitudes against the effective temperature $\teff$ are shown in the bottom panel, representing an asteroseismic HR diagram for our sample of stars. 1$\sigma$ error bars are shown on both quantities for all the plots. The Sun is shown with its usual symbol ($\sun$).}
\label{fig:sample}
\end{center}
\end{figure}

By cross-matching the stars of our sample with the temperature estimates provided by \cite{P12}, we arrived at a final sample of 1640 stars with an accurate $\teff$ (1111 observed in LC and 529 in SC), which will be used for our investigation. Total uncertainties on temperature, as derived by \cite{P12}, include both random and systematic contributions. The amplitudes of the final sample are plotted against $\nmax$ and $\Dnu$ in in Figure~\ref{fig:sample} (top and middle panels, respectively). The bottom panel shows an asteroseismic HR diagram of our sample of stars \citep[amplitudes against $\teff$ from][]{P12}, similar to the one introduced by \cite{Stello10}.  $1\sigma$ error bars are overlaid on both quantities for each panel. The average relative uncertainty in amplitude is $\langle \sigma_\mathrm{A} / A \rangle = 9.2$\,\% for the entire sample, and $\langle \sigma_\mathrm{A} / A \rangle_\mathrm{SC} = 11.7$\,\% and $\langle \sigma_\mathrm{A} / A \rangle_\mathrm{LC} = 8.1$\,\%, for SC and LC targets, respectively.

\section{Amplitude scaling relations}
\label{sec:amp}
Several scaling relations for oscillation amplitudes have been proposed so far. We will briefly introduce them in the following (see \citealt{Huber11} for further discussion).

\subsection{The $L/M$ scaling relation}
The first scaling relation for amplitudes was introduced by \cite{KB95} for radial velocities, based on theoretical models by \cite{CD83}. It is given by
\begin{equation}
v_\mathrm{osc} \propto \left( \frac{L}{M} \right)^s \, ,
\label{eq:lm_osc}
\end{equation}
where $v_\mathrm{osc}$ represents the prediction for the amplitude in radial velocity, $L$ is the luminosity and $M$ the mass of the star, and $s = 1$. \cite{KB95} also showed that the corresponding photometric amplitude $A_{\lambda}$, observed at a wavelength $\lambda$, is related to $v_\mathrm{osc}$ by
\begin{equation}
A_\lambda \propto \frac{v_\mathrm{osc}}{\lambda \teff^{r}} \, .
\label{eq:phot_vel}
\end{equation}
For adiabatic oscillations, the exponent $r$ is $1.5$ \citep[see also][]{Houdek06}, but \cite{KB95} found the observed value for classical p-mode pulsators to be $2.0$ \citep[see also][]{Samadi07}. By combining the two equations, one obtains
\begin{equation}
A_\lambda \propto \left( \frac{L}{M} \right)^s \frac{1}{\lambda \teff^{r }} \, .
\label{eq:lm_prop}
\end{equation}
We are interested in a sample of stars observed with \textit{Kepler}, whose bandpass has a central wavelength $\lambda = 650 \,$nm. By scaling Eq.~(\ref{eq:lm_prop}) to our Sun, we have
\begin{equation}
\frac{A_\lambda}{A_{\lambda,\odot}} = \left( \frac{L/L_{\odot}}{M/M_{\odot}} \right)^s \left(\frac{\teff}{\teffsun} \right)^{-r} \, ,
\label{eq:lm_scaled}
\end{equation}
where $A_\mathrm{650,\odot} = 3.98\,$ppm is the Sun's photometric amplitude observed at the \textit{Kepler} wavelength \citep[e.\,g. see][]{Stello11} and $\teffsun = 5777$\,K. The exponent $s$ has been examined both theoretically \citep[e.\,g. see][]{Houdek99,Houdek02,Samadi07} and observationally \citep{Gilliland08,Dz10,Stello10,Verner11,Baudin11}, and found to be $0.7 < s < 1.5$. The exponent $r$ has usually been chosen to be 2.0 \citep[e.\,g. see the discussion by][]{Stello11}, which implies that solar-like oscillations are not fully adiabatic. This was also shown in the case of \textit{CoRoT} red giant stars by \cite{Samadi12}. Nonetheless, some authors \citep[e.\,g. see][]{Michel08,Mosser10} have chosen to adopt $r=1.5$ \citep[see also the discussion by][]{KB11}.

We can also use the results introduced by \cite{Brown91},
which suggest that the frequency of maximum power scales with the cut-off frequency of the star. Hence, $\nmax \propto g \, / \sqrt{\teff}$, and by considering that $L/M$ scales as $\teff^4 / g$, one obtains
\begin{equation}
\frac{L}{M} \propto \frac{\teff^{3.5}}{\nmax} \,.
\label{eq:lm_teffnu}
\end{equation}
Thus, Eq.~(\ref{eq:lm_scaled}) can be rewritten as 
\begin{equation}
\frac{A_\lambda}{A_{\lambda,\odot}} = \left( \frac{\nmax}{\nmaxsun} \right)^{-s} \left(\frac{\teff}{\teffsun} \right)^{3.5s-r} \, ,
\label{eq:model1}
\end{equation}
where $\nmaxsun = 3100 \, \muhz$. The functional form of the amplitude scaling relation given by Eq.~(\ref{eq:model1}) has the advantage of  simplifying the inference problem presented in Section~\ref{sec:inference} with respect to the one of Eq~(\ref{eq:lm_scaled}). This is because Eq.~(\ref{eq:model1}) is based on a set of independent observables, namely $\nmax$ and $\teff$, which represent the input data used for this work. An analogous argument has been applied to the other scaling relations described in the following sections.

Thus, Eq.~(\ref{eq:model1}) represents the first model to be investigated, which we will refer to as $\mathcal{M}_1$. For this model, both the exponents $s$ and $r$ are set to be free parameters. In parallel, an extended version of Eq.~(\ref{eq:model1}) given by
\begin{equation}
\frac{A_\lambda}{A_{\lambda,\odot}} = \beta \left( \frac{\nmax}{\nmaxsun} \right)^{-s} \left(\frac{\teff}{\teffsun} \right)^{3.5s-r} \, ,
\label{eq:model1off}
\end{equation}
is also considered, where the factor $\beta$ allows the model not to necessarily pass through the solar point, as we will discuss in more detail in Section~\ref{sec:model1}. Eq.~(\ref{eq:model1off}) is treated as a separate model and will be denoted as $\mathcal{M}_{1,\beta}$. Clearly, $\mathcal{M}_{1,\beta}$ depends upon the additional free parameter, $\beta$.

\subsection{The bolometric amplitude}
The stars considered span over a wide range of temperatures, from about $4000$\,K to more than $7000$\,K. Hence, following the discussion by \cite{Huber11}, a more valuable expression for the photometric amplitude could be represented by the bolometric amplitude $A_\mathrm{bol}$, which is related to $A_\lambda$ by \cite[see][]{KB95}:
\begin{equation}
A_\mathrm{bol} \propto \lambda \, A_\lambda \, \teff \propto \frac{v_\mathrm{osc}}{\teff^{r-1}}\, .
\label{eq:bol_lambda}
\end{equation}
By using Eq~(\ref{eq:lm_osc}) we thus have
\begin{equation}
A^{(1)}_\mathrm{bol} \propto \left( \frac{L}{M} \right)^s \frac{1}{\teff^{r -1}} \, ,
\label{eq:amp_bol}
\end{equation}
which by scaling to the Sun and adopting Eq.~(\ref{eq:lm_teffnu}) yields
\begin{equation}
\frac{A_\mathrm{bol}^\mathrm{(1)}}{\abolsun} = \left( \frac{\nmax}{\nmaxsun} \right)^{-s} \left(\frac{\teff}{\teffsun} \right)^{3.5s-r+1} \, ,
\label{eq:model2}
\end{equation}
where $\abolsun = 3.6\,$ppm represents the Sun's bolometric amplitude, determined by \cite{Michel09} and also adopted by \cite{Huber11}.
Eq.~(\ref{eq:model2}) is the second model, $\mathcal{M}_2$, to be investigated in Section~\ref{sec:inference}, with the exponents $s$ and $r$ set again to be the free parameters. We also consider the new model $\mathcal{M}_{2,\beta}$, which again includes the proportionality term $\beta$ playing the same role as in Eq.~(\ref{eq:model1off}).

\subsection{The \textit{Kepler} bandpass-corrected amplitude}
\label{sec:bol_corr}
\cite{Ballot11} have recently established a bolometric correction for amplitude of radial modes observed with \textit{Kepler}, which translates into a correction factor for effective temperatures falling within the range $4000$--$7500$\,K. Again following the approach by \cite{Huber11}, we consider a revised version of Eq.~(\ref{eq:amp_bol}), which reads
\begin{equation}
A^{(2)}_\mathrm{bol} \propto \left( \frac{L}{M} \right)^s \frac{1}{\teff^{r -1} c_K (\teff)} \, ,
\label{eq:amp_bol_corr}
\end{equation}
where
\begin{equation}
c_\mathrm{K} (\teff) = \left( \frac{\teff}{5934} \right)^{0.8}
\label{eq:corr_coeff}
\end{equation}
is the bolometric correction expressed as a power law of the effective temperature. 
By scaling once again to the Sun and applying Eq.~(\ref{eq:lm_teffnu}), we obtain
\begin{equation}
\frac{A^{(2)}_\mathrm{bol}}{\abolsun} = \left( \frac{\nmax}{\nmaxsun} \right)^{-s} \left(\frac{\teff}{\teffsun} \right)^{3.5s-r+0.2} \, ,
\label{eq:model3}
\end{equation}
which we will refer to as $\mathcal{M}_3$. As for the other models, we introduce the model $\mathcal{M}_{3,\beta}$ with the proportionality term $\beta$ included.

\subsection{The mass-dependent scaling relation}
\label{sec:mass_amp}
A mass dependence of the oscillation amplitudes was suggested for the first time by \cite{Huber10redgiant}, and later on studied in detail by \cite{Stello11} for cluster RGs with the introduction of a new scaling relation. It was also tested afterwards by \cite{Huber11} for a wider sample of field stars. According to \cite{Huber11}, an obvious way to modify Eq.~(\ref{eq:amp_bol_corr}) is given by
\begin{equation}
A^{(3)}_\mathrm{bol} \propto \frac{L^s}{M^t} \frac{1}{\teff^{r -1} c_K (\teff)} \, ,
\label{eq:bol_mass}
\end{equation} 
where now the mass varies with the independent exponent $t$. For this case, the dependence upon the quantities $\nmax$ and $\teff$ becomes slightly more complicated because the simple proportionality expressed by Eq.~(\ref{eq:lm_teffnu}) can no longer be adopted. Therefore, the first step to derive the functional form based on our set of observables ($\nmax, \Dnu$, $\teff$), is to consider the scaling relations for the large frequency separation $\Dnu$ \citep[e.\,g. see][]{Bedding07}
\begin{equation}
\frac{\Dnu}{\Dnusun} = \left( \frac{M}{M_\odot} \right)^{0.5} \left( \frac{R}{R_\odot} \right)^{-1.5} \, ,
\label{eq:dnu_scal}
\end{equation}
with $\Dnusun = 135\, \muhz$, and for the frequency of maximum power $\nmax$
\begin{equation}
\frac{\nmax}{\nmaxsun} = \left( \frac{M}{M_\odot} \right) \left( \frac{R}{R_\odot} \right)^{-2} \left( \frac{\teff}{\teffsun} \right)^{-0.5} \, ,
\label{eq:nmax_scal}
\end{equation}
both expressed in terms of the fundamental stellar properties $M$, $R$, and $\teff$. It is however worth mentioning that these scaling relations are empirical approximations whose validity and limitations are not yet fully understood and currently under debate \citep[e.\,g. see][and references therein]{Miglio12}. 
By combining Eq.~(\ref{eq:dnu_scal}) and (\ref{eq:nmax_scal}), one can derive an expression for the seismic radius of a star \citep[e.\,g. see][]{Chaplin11sc}, namely
\begin{equation}
\frac{R}{R_\odot} = \left( \frac{\nmax}{\nmaxsun} \right) \left( \frac{\Dnu}{\Dnusun} \right)^{-2} \left( \frac{\teff}{\teffsun} \right)^{0.5} \, .
\label{eq:radius}
\end{equation}
As a second step, we express $L^s/M^t$ in terms of $R$, $\nmax$, and $\teff$ and scale to the Sun's values, yielding
\begin{equation}
\begin{split}
\left( \frac{L}{L_\odot} \right)^s \left( \frac{M}{M_\odot} \right)^{-t} =&\left( \frac{R}{R_\odot} \right)^{2s-2t} \left(\frac{\nmax}{\nmaxsun} \right)^{-t} \\
&\left(\frac{\teff}{\teffsun} \right)^{4s - 0.5t} \, .
\label{eq:ls_mt}
\end{split}
\end{equation}
Finally, by combining Eqs.~(\ref{eq:corr_coeff}), (\ref{eq:bol_mass}), (\ref{eq:radius}) and (\ref{eq:ls_mt}) we arrive at
\begin{equation}
\begin{split}
\frac{A_\mathrm{bol}^\mathrm{(3)}}{\abolsun} =& \left( \frac{\nmax}{\nmaxsun} \right)^{2s-3t} \left( \frac{\Dnu}{\Dnusun} \right)^{4t - 4s}\\
& \left(\frac{\teff}{\teffsun} \right)^{5s-1.5t -r +0.2 } \, .
\label{eq:model4}
\end{split}
\end{equation}
This represents the model for the mass-dependent scaling relation for amplitudes, hereafter denoted as $\mathcal{M}_4$. In this case, we have the three free parameters $s$, $r$, and $t$ and the set of observables now includes also our measurements of $\Dnu$. The corresponding model $\mathcal{M}_{4,\beta}$ has the largest number of free parameters among those investigated in this work. Note that models $\mathcal{M}_4$ and $\mathcal{M}_{4,\beta}$ reduce to models $\mathcal{M}_3$ and $\mathcal{M}_{3,\beta}$, respectively, for $t = s$.

\subsection{The lifetime-dependent scaling relation}
\label{sec:amp_m5}
\cite{KB11} have recently provided physical arguments to propose a new scaling relation for predicting the amplitudes of solar-like oscillations observed in radial velocities. Their relation arises by postulating that the amplitudes depend on both the stochastic excitation \citep[given by the granulation power, see][for details]{KB11} and the damping rate (given by the mode lifetime). It reads
\begin{equation}
v_\mathrm{osc} \propto \frac{L\tau_\mathrm{osc}^{0.5}}{M^{1.5}\teff^{2.25}} \, ,
\end{equation}
where $\tau_\mathrm{osc}$ is the average mode lifetime of radial modes. By means of Eq.~(\ref{eq:bol_lambda}), and with the bolometric correction introduced in Section~\ref{sec:bol_corr}, the corresponding relation for the bolometric amplitude is given by \citep[see also][]{Huber11}
\begin{equation}
A^{(4)}_\mathrm{bol} \propto \frac{L \tau_\mathrm{osc}^{0.5}}{M^{1.5}} \frac{1}{\teff^{1.25 + r} c_K (\teff)} \, .
\label{eq:amp_m5}
\end{equation} 
In order to obtain the expression for the model $\mathcal{M}_5$, we use similar arguments to those adopted in Section~\ref{sec:mass_amp}, arriving at
\begin{equation}
\begin{split}
\frac{A_\mathrm{bol}^\mathrm{(4)}}{\abolsun} =& \left( \frac{\nmax}{\nmaxsun} \right)^{-2.5} \left( \frac{\Dnu}{\Dnusun} \right)^2 \left( \frac{\tauosc}{\tauoscsun} \right)^{0.5} \\ 
& \left(\frac{\teff}{\teffsun} \right)^{2.3-r} \, ,
\label{eq:model5}
\end{split}
\end{equation}
where $\tauoscsun = 2.88 \,$d, as adopted by \cite{KB11}. For our computations we assume that the mode lifetime is a function of the effective temperature of the star alone \citep[e.\,g. see][]{Chaplin09,Baudin11,App12,Belkacem2012,Corsaro12b}. We used the empirical law found by \cite{Corsaro12b}, which relates the mode linewidths $\Gamma$ of the radial modes ($\ell = 0$) to the effective temperature of the star within the range $4000$--$7000$\,K. In particular, they found that
\begin{equation}
\Gamma = \Gamma_0 \exp{ \left( \frac{\teff - \teffsun}{T_0} \right)} \, ,
\end{equation}
where $\Gamma_0 = 1.39 \pm 0.10\,\muhz$ and $T_0 = 601 \pm 3$\,K. This relation was calibrated using \textit{Kepler} RGs in the open clusters NGC~6791 and NGC~6819, and a sample of MS and subgiant \textit{Kepler} field stars. Given $\tau = (\pi \Gamma)^{-1}$, we obtain
\begin{equation}
\tau_\mathrm{osc} = \tau_0 \exp \left( \frac{\teffsun - \teff}{T_0} \right) \, ,
\label{eq:tau}
\end{equation}
with $\tau_0 = 2.65 \pm 0.19\,$d. The mode lifetimes were computed for all the stars of our sample by means of Eq.~(\ref{eq:tau}), together with their corresponding uncertainties from the error propagation. As for the other scaling relations, we also introduce model $\mathcal{M}_{5,\beta}$. 

\subsection{A new scaling relation}
\label{sec:amp_m6}
Following similar arguments to those adopted by \cite{Stello11} for introducing a new scaling relation for amplitudes of cluster RGs, and the discussion by \cite{Huber11} about the mass dependence of the amplitudes, we introduce a slightly modified version of the amplitude relation given by Eq.~(\ref{eq:amp_m5}), where we set the mass to vary with an independent exponent $t$, thus yielding
\begin{equation}
A^{(5)}_\mathrm{bol} \propto \frac{L \tau_\mathrm{osc}^{0.5}}{M^{t}} \frac{1}{\teff^{1.25 + r} c_K (\teff)} \, .
\end{equation}
By adopting again Eq.~(\ref{eq:ls_mt}) and rearranging, we finally obtain
\begin{equation}
\begin{split}
\frac{A_\mathrm{bol}^\mathrm{(5)}}{\abolsun} =& \left( \frac{\nmax}{\nmaxsun} \right)^{2 -3t} \left( \frac{\Dnu}{\Dnusun} \right)^{4t -4} \left( \frac{\tauosc}{\tauoscsun} \right)^{0.5} \\ 
& \left(\frac{\teff}{\teffsun} \right)^{4.55-r-1.5t} \, ,
\label{eq:model6} 
\end{split}
\end{equation}
hereafter marked as model $\mathcal{M}_6$. As done for the other scaling relations, the model $\mathcal{M}_{6, \beta}$ is also included in our inference. Clearly, models $\mathcal{M}_6$ and $\mathcal{M}_{6, \beta}$ reduce to models $\mathcal{M}_5$ and $\mathcal{M}_{5, \beta}$, respectively, for $t = 1.5$.

\section{Bayesian inference}
\label{sec:inference}
We now use Bayesian inference for the free parameters of the models described above. The Bayes' theorem tells us that
\begin{equation}
p ( \boldsymbol{\xi} \mid A, \mathcal{M}) = \frac{ p( A \mid \boldsymbol{\xi}, \mathcal{M} ) \pi ( \boldsymbol{\xi} \mid \mathcal{M})}{p (A \mid \mathcal{M})} \, ,
\label{eq:bayes}
\end{equation}
where $\boldsymbol{\xi} = \xi_1, \xi_2, \dots, \xi_k$ is the vector of the $k$ free parameters that formalize the hypotheses of the model $\mathcal{M}$, considered for the inference, and $A$ is the set of amplitude measurements. The term $p( A \mid \boldsymbol{\xi}, \mathcal{M} )$ is now identified with the likelihood $\mathcal{L}(\boldsymbol{\xi})$ of the parameters $\xi_i$ given the measured oscillation amplitudes:
\begin{equation}
p(A \mid \boldsymbol{\xi}, \mathcal{M}) = \mathcal{L} (\boldsymbol{\xi} \mid A, \mathcal{M}) \, .
\end{equation}
Thus, the left-hand side of Eq.~(\ref{eq:bayes}) is the posterior probability density function (PDF), while the right-hand side is the product of the likelihood function $\mathcal{L} (\boldsymbol{\xi})$, which represents our manner of comparing the data to the predictions by the model, and the prior PDF $\pi(\boldsymbol{\xi} \mid \mathcal{M})$, which represents our knowledge of the inferred parameters before any information from the data is available. The term $p(A \mid \mathcal{M})$ is a normalization factor, known as the Bayesian evidence, which we do not consider for the inference problem because it is a constant for a model alone. As we will argue in Section~\ref{sec:evidence}, the Bayesian evidence is essential for solving the problem of model comparison.

For our inference problem, we adopt the common Gaussian likelihood function, which presumes that the residuals arising from the difference between observed and predicted logarithms of the amplitudes are Gaussian distributed, i.\,e. the amplitudes themselves are presumed to be log-normal distributed \citep[see also][]{App98}. Therefore, we have
\begin{equation}
\mathcal{L} (\boldsymbol{\xi}) = \prod^{N}_{i=1} \frac{1}{\sqrt{2\pi} \widetilde{\sigma}_i} \exp {\left[ - \frac{1}{2} \left( \frac{ \Delta_i (\boldsymbol{\xi})}{\widetilde{\sigma}_i} \right)^2 \right]} \, ,
\label{eq:likelihood}
\end{equation} 
where $N$ is the total number of data points (the number of stars, in our case), while
\begin{equation}
\Delta_i (\boldsymbol{\xi}) = \ln A^\mathrm{obs}_i - \ln A^\mathrm{th}_i (\boldsymbol{\xi}) 
\label{eq:delta}
\end{equation}
is the difference between the observed logarithmic amplitude for the $i$-th star and the predicted one (which depends on the adopted model, i.\,e. on the parameters vector $\boldsymbol{\xi}$). The term $\widetilde{\sigma}_i$ appearing in the leading exponential term of Eq.~(\ref{eq:likelihood}) is the total uncertainty in the predicted logarithmic amplitude, namely the relative uncertainty of the amplitude enlarged by the relative errors of the independent variables $\nmax$, $\Dnu$, and $\teff$. This means that we are not considering error-free variables in our models \citep[see e.\,g.][for more details]{error2,error1}. For simplifying the computations a modified version of the likelihood function, known as the \textit{log-likelihood}, is preferred. The log-likelihood function is defined as $\Lambda (\boldsymbol{\xi}) \equiv \ln \mathcal{L} (\boldsymbol{\xi})$, which yields
\begin{equation}
\Lambda (\boldsymbol{\xi}) = \Lambda_0 - \frac{1}{2} \sum^{N}_{i=1} \left[ \frac{ \Delta_i (\boldsymbol{\xi}) }{\widetilde{\sigma}_i} \right]^2 \, ,
\label{eq:log_likelihood}
\end{equation}
where 
\begin{equation}
\Lambda_0 = - \sum^{N}_{i=1} \ln {\sqrt{2 \pi} \widetilde{\sigma}_i} \, .
\end{equation}
The choice of reliable priors is important in the Bayesian approach. For our purpose, uniform priors represent a sensible choice for most of the free parameters. This means one has no assumptions about the inferred parameters before any knowledge coming from the data, with equal weight being given to all values of each of the parameters considered. In particular, we use standard uniform priors for the exponents $s$, $r$ and $t$ of the models described above, letting the parameters vary within a limited range in order to make the priors proper, i.\,e. normalizable to unity. For the proportionality term $\beta$ introduced in Eq.~(\ref{eq:model1off}), we adopt the Jeffreys' prior $\propto \beta^{-1}$ \citep{Kass}, a class of uninformative prior that results in a uniform prior for the natural logarithm of the parameter. In this manner, the parameter of interest is represented by the offset $\ln \beta$ (see below), whose prior is uniform distributed and also limited in range. Hence, uniform priors are included in the inference problem as a simple constant term depending on the intervals adopted for the inferred parameters
\begin{equation}
\pi ( \boldsymbol{\xi} \mid \mathcal{M}) = \prod^4_{j=1} \left[ \xi^\mathrm{max}_j - \xi^\mathrm{min}_j \right]^{-1} \,
\label{eq:prior_tot}
\end{equation}
with $\xi_1 = s$, $\xi_2 = r$, $\xi_3 = t$ and $\xi_4 = \ln \beta$, and $\xi^\mathrm{min}_j, \xi^\mathrm{max}_j$ the minimum and maximum values defining the interval of the $j$-th parameter. The intervals that we adopt are listed in Table~\ref{tab:intervals_amp}. These ranges are used for both the Bayesian parameter estimation and the model comparison.
\begin{table}
\centering
\footnotesize
 \caption{Maximum and minimum values of the free parameters, adopted for all the models and samples.}
\begin{tabular}{lr}
  \hline
  \\[-8pt]
 $\xi_j$& \multicolumn{1}{c}{$\left[ \xi^\mathrm{min}_j, \xi^\mathrm{max}_j \right]$}\\[2pt]
  \hline
  \\[-8pt]
  $s$ & $\left[ 0.2, 1.2 \right]$\\
  $r$ & $\left[ -6.5, 11.0 \right]$ \\
  $t$ & $\left[ 1.0, 2.0 \right]$ \\
  $\ln \beta$ & $\left[ -1.0, 1.0\right]$ \\[2pt]
  \hline
 \end{tabular}
\label{tab:intervals_amp}
\end{table}

A note of caution concerns the treatment of the uncertainties. In fact, by using the natural logarithm of the equations that describe the models (see also Section~\ref{sec:model1}), we ensure that we are not favoring for example frequencies upon periods, amplitudes upon power, temperatures upon surface brightness, etc., which has the advantage of making the error propagation law fully correct. Thus, the corresponding uncertainties to be considered in Eq.~(\ref{eq:likelihood}) will be the relative uncertainties.

The inference problem for a given parameter, e.\,g. $\xi_1$, is then performed by integrating (marginalizing) the posterior distribution function $p ( \boldsymbol{\xi} \mid d, \mathcal{M})$ over the remaining $k-1$ parameters $\xi_2, \xi_3, \dots, \xi_k$. We obtain the corresponding marginal PDF of the parameter $\xi_1$
\begin{equation}
p ( \xi_1 \mid A, \mathcal{M}) = \int p ( \boldsymbol{\xi} \mid A, \mathcal{M}) d\xi_2, d\xi_3, \dots, d\xi_k \, ,
\label{eq:marginal}
\end{equation}
whose statistical moments and credible intervals (i.\,e. Bayesian confidence intervals) are the quantities of interest for our work. Since the dimensionality of our problem is not higher than $k = 4$, all the integrations can be computed by direct numerical summation of the posterior distribution over the remaining parameters (only in case the observables, like $\teff$, were error-free all integrations could be computed analytically).

The results presented in the coming sections are derived in three cases: for the entire sample of stars, and for SC (dominated by MS stars) and LC (RGs) targets separately. Analyzing the two subsets separately allows us to test whether the fitted parameters of the scaling relations are sensitive to the evolutionary stage of the stars \citep[see also][]{Huber11}.

The mean values (or expectation values) of the free parameters of the models, together with their corresponding 68.3\,\% Bayesian credible intervals, are listed in Table~\ref{tab:model} for the case of the entire sample, and in Tables~\ref{tab:modelsc} and~\ref{tab:modellc} for the subsets of SC and LC targets. We also computed a weighted Gaussian rms, $\sigma^w_\mathrm{rms}$, of the residuals $\Delta^2_i (\boldsymbol{\xi})$, where we adopted the weights $w_i = \widetilde{\sigma}^{-2}_i$, $\widetilde{\sigma_i}$ being the total uncertainty used in Eq.~(\ref{eq:likelihood}). The maximum of the log-likelihood function, $\Lambda_\mathrm{max}$, and $\sigma^w_\mathrm{rms}$, used as an estimate of the fit quality, are also listed in the same tables. In addition, we derived correlation coefficients for each pair of free parameters by means of principal component analysis using Singular Value Decomposition (SVD) from the posterior PDFs. The results are shown in Tables~\ref{tab:correlation},~\ref{tab:correlationsc},~\ref{tab:correlationlc} for the entire sample, and SC and LC targets, respectively, with $-1$ meaning total anti-correlation and $1$ total correlation. The effects of the correlations will be discussed in Section~\ref{sec:amp_discussion}.

\subsection{Statistically independent models}
Before getting to a description on how to correctly include the models in the inference problem, it is useful to highlight that the models $\mathcal{M}_{1,\beta}$, $\mathcal{M}_{2,\beta}$, and $\mathcal{M}_{3,\beta}$ on the one hand, and the models $\mathcal{M}_2$, and $\mathcal{M}_3$ the other hand, are statistically identical to one another (but not identical in the general sense, since their underlying physical assumptions are different). In case the intervals of their free parameters are the same for all the models when performing the Bayesian parameter estimation (as it is for the analysis presented in this work) this implies that a statistical inference for these models would lead to identical (or directly related) values of these free parameters. In particular, according to Eqs.~(\ref{eq:model1off}),~(\ref{eq:model2}), and (\ref{eq:model3}) we have that 
\begin{equation}
\begin{split}
s(\mathcal{M}_{1,\beta}) &= s(\mathcal{M}_{2,\beta}) = s(\mathcal{M}_{3,\beta})\\
s(\mathcal{M}_2) &= s(\mathcal{M}_3)
\label{eq:rel_s}
\end{split}
\end{equation}
for the exponent $s$,
\begin{equation}
\begin{split}
r(\mathcal{M}_{1,\beta}) &= r(\mathcal{M}_{2,\beta}) - 1 = r(\mathcal{M}_{3,\beta}) - 0.2\\
r(\mathcal{M}_2) &= r(\mathcal{M}_3) - 0.8 \, ,
\label{eq:rel_r}
\end{split}
\end{equation}
for the exponent $r$, and
\begin{equation}
\begin{split}
\ln \beta(\mathcal{M}_{1,\beta}) &= \ln \beta(\mathcal{M}_{2,\beta}) + \ln \left(\frac{\abolsun}{A_{\lambda,\odot}} \right) \\
&= \ln \beta(\mathcal{M}_{3,\beta}) + \ln \left(\frac{\abolsun}{A_{\lambda,\odot}} \right) \\
\label{eq:rel_offset}
\end{split}
\end{equation}
for the offset $\ln \beta$, where the term $\ln \left(\abolsun / A_{\lambda,\odot} \right)$ arises from the difference in considering the amplitudes to be either observed at $\lambda = 650\,$nm ($\mathcal{M}_{1,\beta}$) or bolometric ($\mathcal{M}_{2,\beta}$ and $\mathcal{M}_{3,\beta}$).  As a consequence, from now on the entire analysis for the models $\mathcal{M}_{1,\beta}$, $\mathcal{M}_{2,\beta}$, and $\mathcal{M}_{3,\beta}$ on one side, and for the models $\mathcal{M}_2$ and $\mathcal{M}_3$ on the other side, will be reduced to that of the two models $\mathcal{M}_{1,\beta}$ and $\mathcal{M}_2$, respectively. The reader can derive the corresponding parameters for the other dependent models using Eqs.~(\ref{eq:rel_s}), (\ref{eq:rel_r}), and (\ref{eq:rel_offset}).

\begin{table*}
\begin{minipage}{185mm}
\centering
 \caption{Expectation values of the inferred parameters for all the models described in Section~\ref{sec:amp} in the case of the entire sample (both LC and SC targets), having $N = 1640$ stars. 68.3\,\% Bayesian credible intervals are added. The maximum value for the log-likelihood function $\Lambda_\mathrm{max}$ and a weighted Gaussian rms of the residuals, $\sigma^w_\mathrm{rms}$, are also reported.}
\begin{tabular}{llrcrrc}
  \hline
  \\[-8pt]
 Model & \multicolumn{1}{c}{$s$} & \multicolumn{1}{c}{$r$} & \multicolumn{1}{c}{$t$} & \multicolumn{1}{c}{$\ln \beta$} & \multicolumn{1}{c}{$\Lambda_\mathrm{max}$} & $\sigma^w_\mathrm{rms}\,$\\[1pt]
  \hline
  \\[-8pt]
  $\mathcal{M}_1$ & $0.680 \pm 0.002 $ & $4.31 \pm 0.04$ & -- & \multicolumn{1}{c}{--} & $-3533.1$ & 0.23\\[1pt]
  $\mathcal{M}_{1,\beta}$ & $0.524 \pm 0.004$ & $5.51 \pm 0.05$ & \multicolumn{1}{c}{--} & $0.400 \pm 0.010$ & $-2491.3$ & 0.24 \\[1pt]
  $\mathcal{M}_2$ & $0.722 \pm 0.002$ & $4.96 \pm 0.04$ & -- & \multicolumn{1}{c}{--} & $-4159.9$ & 0.24 \\[1pt]
  $\mathcal{M}_4$ & $0.822 \pm 0.003$ & $3.93 \pm 0.06$ & $1.58 \pm 0.02$ & \multicolumn{1}{c}{--} & $-948.5$ & 0.23 \\[1pt]
  $\mathcal{M}_{4,\beta}$ & $0.643 \pm 0.005$ & $4.46 \pm 0.06$ & $1.36 \pm 0.02$ & $0.528 \pm 0.012$ & 163.4 & 0.18\\[1pt]
  $\mathcal{M}_5$ & \multicolumn{1}{c}{--} & $-5.71 \pm 0.03$ & -- & \multicolumn{1}{c}{--} & $-925.5$ & 0.31 \\[1pt]
  $\mathcal{M}_{5,\beta}$ & \multicolumn{1}{c}{--} & $-5.07 \pm 0.05$ & -- & $-0.122 \pm 0.007$ & $-786.9$ & 0.29 \\[1pt]
  $\mathcal{M}_6$ & \multicolumn{1}{c}{--} & $-5.04 \pm 0.05$ & $1.80 \pm 0.02$ & \multicolumn{1}{c}{--} & $-727.1$ & 0.30\\[1pt]
  $\mathcal{M}_{6,\beta}$ & \multicolumn{1}{c}{--} & $-4.97 \pm 0.06$ & $1.75 \pm 0.02$ & $-0.035 \pm 0.012$ & $-722.8$ & 0.29\\[2pt]
  \hline
 \end{tabular}
\label{tab:model}
\end{minipage}
\end{table*}
\begin{table*}
\begin{minipage}{185mm}
\centering
 \caption{Same description as for Table~\ref{tab:model} but in the case of SC targets, having $N = 529$ stars.}
 \begin{tabular}{llrcrrc}
  \hline
  \\[-8pt]
  Model & \multicolumn{1}{c}{$s$} & \multicolumn{1}{c}{$r$} & \multicolumn{1}{c}{$t$} & \multicolumn{1}{c}{$\ln \beta$} & \multicolumn{1}{c}{$\Lambda_\mathrm{max}$} & $\sigma^w_\mathrm{rms}\,$\\[1pt]
  \hline
  \\[-8pt]
  $\mathcal{M}_1$ & $0.775 \pm 0.003 $ & $3.23 \pm 0.05$ & -- & \multicolumn{1}{c}{--} & $-321.4$ & 0.20\\[1pt]
  $\mathcal{M}_{1,\beta}$ & $0.624 \pm 0.010$ & $3.68 \pm 0.06$ & -- & $0.241 \pm 0.016$ & $-199.3$ & 0.19 \\[1pt]
  $\mathcal{M}_2$ & $0.838 \pm 0.003$ & $4.05 \pm 0.05$ & -- & \multicolumn{1}{c}{--} & $-443.8$ & 0.22\\[1pt]
  $\mathcal{M}_4$ & $0.984^{+0.009}_{-0.010}$ & $2.79 \pm 0.09$ & $1.66^{+0.04}_{-0.05}$ & \multicolumn{1}{c}{--} & $-94.9$ & 0.23 \\[1pt]
  $\mathcal{M}_{4,\beta}$ & $0.748 \pm  0.015$ & $3.47 \pm 0.09$ & $1.27 \pm 0.04$ & $0.321 \pm 0.020$ & 18.0 & 0.20 \\[1pt]
  $\mathcal{M}_5$ & \multicolumn{1}{c}{--} & $-2.78 \pm 0.09$ & -- & \multicolumn{1}{c}{--} & $-83.1$ & 0.24 \\[1pt]
  $\mathcal{M}_{5,\beta}$ & \multicolumn{1}{c}{--} & $-2.75 \pm 0.09$ & -- & $0.020 \pm 0.008$ & $-80.2$ & 0.24 \\[1pt]
  $\mathcal{M}_6$ & \multicolumn{1}{c}{--} & $-2.80^{+0.10}_{-0.09}$ & $1.56 \pm 0.02$ & \multicolumn{1}{c}{--} & $-79.6$ & 0.24 \\[1pt]
  $\mathcal{M}_{6,\beta}$ & \multicolumn{1}{c}{--} & $-2.75 \pm 0.10$ & $1.72 \pm 0.04$ & $0.087^{+0.014}_{-0.015}$ & $-59.4$ & 0.24\\[2pt]
  \hline
 \end{tabular}
\label{tab:modelsc}
\end{minipage}
\end{table*}~
\begin{table*}
\begin{minipage}{185mm}
\centering
 \caption{Same description as for Table~\ref{tab:model} but in the case of LC targets, having $N = 1111$ stars.}
 \begin{tabular}{llrcrrc}
  \hline
  \\[-8pt]
  Model & \multicolumn{1}{c}{$s$} & \multicolumn{1}{c}{$r$} & \multicolumn{1}{c}{$t$} & \multicolumn{1}{c}{$\ln \beta$} & \multicolumn{1}{c}{$\Lambda_\mathrm{max}$} & $\sigma^w_\mathrm{rms}\,$\\[1pt]
  \hline
  \\[-8pt]
  $\mathcal{M}_1$ & $0.464^{+0.007}_{-0.006}$ & $9.53^{+0.15}_{-0.16}$ & -- & -- & $-799.2$ & 0.28\\[1pt]
  $\mathcal{M}_{1,\beta}$ & $0.548 \pm 0.009$ & $9.67^{+0.15}_{-0.16}$ & -- & $-0.35 \pm 0.03$ & $-737.4$ & 0.27 \\[1pt]
  $\mathcal{M}_2$ & $0.491^{+0.007}_{-0.006}$ & $10.51^{+0.15}_{-0.16}$ & -- & -- & $-769.0$ & 0.28\\[1pt]
  $\mathcal{M}_4$ & $0.666^{+0.006}_{-0.005}$ & $6.99 \pm 0.12$ & $1.28 \pm 0.02$ & -- & $207.4$ & 0.18 \\[1pt]
  $\mathcal{M}_{4,\beta}$ & $0.602 \pm 0.008$ & $5.87 \pm  0.14$ & $1.31 \pm 0.02$ & $0.45 \pm 0.03$ & 301.0 & 0.16 \\[1pt]
  $\mathcal{M}_5$ & \multicolumn{1}{c}{--} & $-6.08 \pm 0.04$ & -- & -- & $-357.9$ & 0.24 \\[1pt]
  $\mathcal{M}_{5,\beta}$ & \multicolumn{1}{c}{--} & $-4.38 \pm 0.16$ & -- & $-0.27^{+0.02}_{-0.03}$ & $-309.2$ & 0.24 \\[1pt]
  $\mathcal{M}_6$ & \multicolumn{1}{c}{--} & $-5.94 \pm 0.08$ & $1.55 \pm 0.03$ & -- & $-356.0$ & 0.24\\[1pt]
  $\mathcal{M}_{6,\beta}$ & \multicolumn{1}{c}{--} & $-4.39 \pm 0.16$ & $1.45^{+0.02}_{-0.03}$ & $-0.30 \pm 0.03$ & $-307.1$ & 0.24\\[2pt]
  \hline
 \end{tabular}
\label{tab:modellc}
\end{minipage}
\end{table*}

\begin{table*}
\begin{minipage}{185mm}
\centering
 \caption{Correlation coefficients for pairs of free parameters for each model in the case of the entire sample.}
 \begin{tabular}{lrrrrrr}
  \hline
  \\[-8pt]
  Model & \multicolumn{1}{c}{$s$ vs $r$} & \multicolumn{1}{c}{$s$ vs $b$} & \multicolumn{1}{c}{$s$ vs $t$} & \multicolumn{1}{c}{$r$ vs $b$} & \multicolumn{1}{c}{$r$ vs $t$} & \multicolumn{1}{c}{$b$ vs $t$}\\[1pt]
  \hline
  \\[-8pt]
  $\mathcal{M}_1$ & $-0.90$ & -- & -- & -- & -- & --\\[1pt]
  $\mathcal{M}_{1,\beta}$ & $-0.85$ & $-0.92$ & -- & $0.66$ & -- & -- \\[1pt]
  $\mathcal{M}_2$ & $-0.94$ & -- & -- & -- & -- & --\\[1pt]
  $\mathcal{M}_4$ & $-0.74$ & -- & $0.59$ & -- & $0.01$ & -- \\[1pt]
  $\mathcal{M}_{4,\beta}$ & $-0.74$ & $-0.76$ & $0.44$ & $0.31$ & $-0.25$ & $0.04$ \\[1pt]
  $\mathcal{M}_5$ & -- & -- & -- & -- & -- & -- \\[1pt]
  $\mathcal{M}_{5,\beta}$ & -- & -- & -- & $-0.74$ & -- & -- \\[1pt]
  $\mathcal{M}_6$ & -- & -- & -- & -- & $0.74$ & --\\[1pt]
  $\mathcal{M}_{6,\beta}$ & -- & -- & -- & $-0.41$ & $0.18$ & $0.71$\\[2pt]
  \hline
 \end{tabular}
\label{tab:correlation}
\end{minipage}
\end{table*}
\begin{table*}
\begin{minipage}{185mm}
\centering
 \caption{Same description as for Table~\ref{tab:correlation} but in the case of SC targets.}
 \begin{tabular}{lrrrrrr}
  \hline
  \\[-8pt]
  Model & \multicolumn{1}{c}{$s$ vs $r$} & \multicolumn{1}{c}{$s$ vs $b$} & \multicolumn{1}{c}{$s$ vs $t$} & \multicolumn{1}{c}{$r$ vs $b$} & \multicolumn{1}{c}{$r$ vs $t$} & \multicolumn{1}{c}{$b$ vs $t$}\\[1pt]
  \hline
  \\[-8pt]
  $\mathcal{M}_1$ & $0.22$ & -- & -- & -- & -- & --\\[1pt]
  $\mathcal{M}_{1,\beta}$ & $-0.40$ & $-0.94$ & -- & $0.49$ & -- & -- \\[1pt]
  $\mathcal{M}_2$ & $0.20$ & -- & -- & -- & -- & --\\[1pt]
  $\mathcal{M}_4$ & $-0.29$ & -- & $0.88$ & -- & $-0.27$ & -- \\[1pt]
  $\mathcal{M}_{4,\beta}$ & $-0.48$ & $-0.85$ & $0.71$ & $0.44$ & $-0.37$ & $-0.31$ \\[1pt]
  $\mathcal{M}_5$ & -- & -- & -- & -- & -- & -- \\[1pt]
  $\mathcal{M}_{5,\beta}$ & -- & -- & -- & $0.12$ & -- & -- \\[1pt]
  $\mathcal{M}_6$ & -- & -- & -- & -- & $-0.07$ & --\\[1pt]
  $\mathcal{M}_{6,\beta}$ & -- & -- & -- & $0.03$ & $-0.01$ & $0.76$\\[2pt]
  \hline
 \end{tabular}
\label{tab:correlationsc}
\end{minipage}
\end{table*}
\begin{table*}
\begin{minipage}{185mm}
\centering
 \caption{Same description as for Table~\ref{tab:correlation} but in the case of LC targets.}
 \begin{tabular}{lrrrrrr}
  \hline
  \\[-8pt]
  Model & \multicolumn{1}{c}{$s$ vs $r$} & \multicolumn{1}{c}{$s$ vs $b$} & \multicolumn{1}{c}{$s$ vs $t$} & \multicolumn{1}{c}{$r$ vs $b$} & \multicolumn{1}{c}{$r$ vs $t$} & \multicolumn{1}{c}{$b$ vs $t$}\\[1pt]
  \hline
  \\[-8pt]
  $\mathcal{M}_1$ & $-0.98$ & -- & -- & -- & -- & --\\[1pt]
  $\mathcal{M}_{1,\beta}$ & $-0.49$ & $-0.71$ & -- & $-0.25$ & -- & -- \\[1pt]
  $\mathcal{M}_2$ & $-0.98$ & -- & -- & -- & -- & --\\[1pt]
  $\mathcal{M}_4$ & $-0.92$ & -- & $0.50$ & -- & $-0.19$ & -- \\[1pt]
  $\mathcal{M}_{4,\beta}$ & $-0.31$ & $-0.66$ & $0.30$ & $-0.46$ & $-0.30$ & $0.18$ \\[1pt]
  $\mathcal{M}_5$ & -- & -- & -- & -- & -- & -- \\[1pt]
  $\mathcal{M}_{5,\beta}$ & -- & -- & -- & $-0.98$ & -- & -- \\[1pt]
  $\mathcal{M}_6$ & -- & -- & -- & -- & $0.88$ & --\\[1pt]
  $\mathcal{M}_{6,\beta}$ & -- & -- & -- & $-0.88$ & $0.04$ & $0.37$\\[2pt]
  \hline
 \end{tabular}
\label{tab:correlationlc}
\end{minipage}
\end{table*}

\subsection{Models $\mathcal{M}_1$ and $\mathcal{M}_{1,\beta}$}
\label{sec:model1}
The first model to be investigated is given by Eq.~(\ref{eq:model1}). As argued above, we need to consider the natural logarithm in order to treat the observables independently of the function adopted (see the discussion in Section~\ref{sec:inference}). Hence the model reads
\begin{equation}
\begin{split}
\ln \left( \frac{A_\lambda}{A_{\lambda,\odot}} \right) = &- s \ln \left( \frac{\nmax}{\nmaxsun} \right) \\
&+ ( 3.5s - r) \ln \left( \frac{\teff}{\teffsun} \right) \, .
\label{eq:lnmodel1}
\end{split}
\end{equation}
At this stage we briefly describe how the uncertainties have been included in our analysis. The new uncertainties on the scaled amplitude to be considered in Eq.~(\ref{eq:likelihood}) are clearly given by $\sigma_{A_i}/A_i$, hereafter $\widetilde{\sigma}_{A_i}$ for simplicity, where $A_i$ is the observed amplitude for the $i$-th star and $\sigma_{A_i}$ its corresponding uncertainty as derived by \cite{Huber11}. 
However, Eq.~(\ref{eq:lnmodel1}) suggests that the uncertainty on amplitude is not the only one affecting the predicted amplitude $A_\lambda$. In fact, uncertainties on both $\nmax$ \citep[derived by][]{Huber11} and $\teff$ \citep[from][]{P12} have to be included in our computations. The total uncertainty to be used in Eq.~(\ref{eq:likelihood}) is given by
the Gaussian error propagation law, which gives
\begin{equation}
\widetilde{\sigma}^2_i (s, r) = \widetilde{\sigma}^2_{A_i} + s^2 \widetilde{\sigma}^2_{\nu_{\mathrm{max},i}} + (3.5s -r)^2 \widetilde{\sigma}^2_{T_{\mathrm{eff},i}}
\label{eq:tot_uncert_model1}
\end{equation}
where we defined $\widetilde{\sigma}_{\nu_{\mathrm{max},i}} \equiv \sigma_{\nu_{\mathrm{max},i}} / \nu_{\mathrm{max},i}$ and $\widetilde{\sigma}_{T_{\mathrm{eff},i}} \equiv \sigma_{T_{\mathrm{eff},i}} / T_{\mathrm{eff},i}$, similarly to what was done for the amplitudes. Eq.~(\ref{eq:tot_uncert_model1}) only holds in case of uncorrelated uncertainties and linear relations \citep[see also][]{error2,error1}. Intuitively, Eq.~(\ref{eq:tot_uncert_model1}) is the quadratic sum of the relative uncertainties over the physical quantities considered, according to Eq.~(\ref{eq:lnmodel1}). 
A variation of Eq.~(\ref{eq:lnmodel1}) is represented by the model $\mathcal{M}_{1,\beta}$ given by Eq.~(\ref{eq:model1off}), whose natural logarithm reads 
\begin{equation}
\begin{split}
\ln \left( \frac{A_\lambda}{A_{\lambda,\odot}} \right) = &- s \ln \left( \frac{\nmax}{\nmaxsun} \right) \\
&+ ( 3.5s - r) \ln \left( \frac{\teff}{\teffsun} \right) + \ln \beta \, .
\label{eq:lnmodel1off}
\end{split}
\end{equation}
which differs from Eq.~(\ref{eq:lnmodel1}) by the additional term $\ln \beta$. As already argued before, the offset $\ln \beta$ allows the model not to necessarily pass through the solar point $(A_\odot, \nmaxsun, \teffsun)$. Its introduction in the inference is of importance if one wants to assess whether or not the Sun is a good reference star for the sample considered. This choice is also motivated by the fact that the Sun is falling at the edge of the sample of stars when plotting amplitudes against $\nmax$ and $\Dnu$ (Figure~\ref{fig:sample}, top and middle panels). This peculiar position is also evident from our asteroseismic HR diagram (Figure~\ref{fig:sample}, bottom panel), and is caused by the lack of solar twins in our sample of stars \citep[see the discussion by][]{Chaplin11sc}. In fact, in the case of $\ln \beta \ne 0$, by replacing the solar values in Eq.~(\ref{eq:lnmodel1off}) (or alternatively Eq.~(\ref{eq:model1off})), the predicted amplitude for the Sun would be resized by a factor $\beta$. This means that the best reference amplitude for scaling the amplitudes of our sample of stars would be represented by $\beta A_{\lambda,\odot}$.  According to Eq.~(\ref{eq:lnmodel1off}), the total uncertainty to be considered in building the likelihood function for the model $\mathcal{M}_{1,\beta}$ is given again by Eq.~(\ref{eq:tot_uncert_model1}) because the offset does not play any role in the total contribute of the uncertainties.

A representative sample of the resulting marginal PDFs is plotted in Figure~\ref{fig:marg_m1} for the case of the entire sample, where 68.3\,\% Bayesian credible regions (shaded bands) and expectation values (dashed lines) are also marked. The comparison between the predicted and the observed amplitudes is shown in Figure~\ref{fig:model1} for the three cases considered (top panels for model $\mathcal{M}_1$, bottom panels for model $\mathcal{M}_{1,\beta}$), together with a plot of the residuals arising from the difference between the models and the observations.

\subsection{Model $\mathcal{M}_2$}
\label{sec:model2}
The model $\mathcal{M}_2$ given by Eq.~(\ref{eq:model2}) deserves a similar description to that presented in Section~(\ref{sec:model1}) for model $\mathcal{M}_1$, where the natural logarithm is now given by
\begin{equation}
\begin{split}
\ln \left( \frac{A^\mathrm{(1)}_\mathrm{bol}}{\abolsun} \right) = &- s \ln \left( \frac{\nmax}{\nmaxsun} \right) \\
&+ ( 3.5s - r + 1) \ln \left( \frac{\teff}{\teffsun} \right) \, ,
\label{eq:lnmodel2}
\end{split}
\end{equation}
with a total uncertainty for the $i$-th star expressed as
\begin{equation}
\widetilde{\sigma}^2_i (s, r) = \widetilde{\sigma}^2_{A_i} + s^2 \widetilde{\sigma}^2_{\nu_{\mathrm{max},i}} + (3.5s -r + 1)^2 \widetilde{\sigma}^2_{T_{\mathrm{eff},i}} \, \,
\label{eq:tot_uncert_model2}
\end{equation}
to be included in Eq.~(\ref{eq:likelihood}). The resulting models are shown in Figure~\ref{fig:model2}, with similar descriptions as those adopted for Figure~\ref{fig:model1}. 

\subsection{Models $\mathcal{M}_4$ and $\mathcal{M}_{4,\beta}$}
\label{sec:model4}
The models $\mathcal{M}_4$ and $\mathcal{M}_{4,\beta}$ (see Section~\ref{sec:mass_amp}) are clearly the most complex ones among those investigated in this work because the largest number of free parameters is involved, and measurements of $\Dnu$ are also needed. We note that, although a tight correlation between $\nmax$ and $\Dnu$ has been found in previous studies \citep[e.\,g. see][]{Stello09a}, we choose not to adopt the $\nmax$--$\Dnu$ relation to express model $\mathcal{M}_4$ in terms of $\nmax$ only (or alternatively $\Dnu$) because additional uncertainties arising from the scatter around this relation would affect the results of our inference. This is also motivated by recent results by \cite{Huber11} who found that the $\nmax$-$\Dnu$ relation changes as a function of $\teff$ between dwarf and giant stars.

Therefore, by considering the natural logarithm of Eq.~(\ref{eq:model4}) one obtains
\begin{equation}
\begin{split}
\ln \left( \frac{A^{(3)}_\mathrm{bol}}{\abolsun} \right) =& \, \, (2s -3t) \ln \left( \frac{\nmax}{\nmaxsun} \right)+ (4t -4s) \ln \left( \frac{\Dnu}{\Dnusun} \right)\\
& + (5s - 1.5t -r + 0.2) \ln \left( \frac{\teff}{\teffsun} \right) \, ,
\label{eq:lnmodel4}
\end{split}
\end{equation}
for model $\mathcal{M}_4$, and with the additional term $\ln \beta$ for $\mathcal{M}_{4,\beta}$. According to Eq.~(\ref{eq:lnmodel4}), the total uncertainty to be considered in Eq.~(\ref{eq:likelihood}) reads
\begin{equation}
\begin{split}
\widetilde{\sigma}^2_i (s, r, t) =&\, \widetilde{\sigma}^2_{A_i} + (2s-3t)^2 \widetilde{\sigma}^2_{\nu_{\mathrm{max},i}} + (4t-4s)^2 \widetilde{\sigma}^2_{\Dnu_i} \\
&+ (5s -1.5t -r + 0.2)^2 \widetilde{\sigma}^2_{T_{\mathrm{eff},i}} \, ,
\label{eq:tot_uncert_model4}
\end{split}
\end{equation}
with $\widetilde{\sigma}_{\Dnu_i} \equiv \sigma_{\Dnu_i} / \Dnu_i$, as done for the other quantities. As one can intuitively expect, the new total uncertainty depends on the three free parameters of the model. The resulting models are shown in Figure~\ref{fig:model4}.

\subsection{Models $\mathcal{M}_5$ and $\mathcal{M}_{5,\beta}$}
The models described in Section~\ref{sec:amp_m5} are derived with a quite different approach, which requires an estimate of the mode lifetime for each star considered in our sample. We note that model $\mathcal{M}_5$ was also investigated by \cite{Huber11}, who however did not take into account mode lifetimes. 

\begin{figure*}
\begin{center}
\includegraphics[scale=0.61]{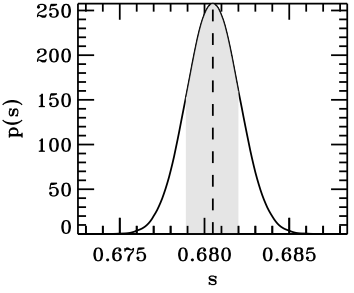}\includegraphics[scale=0.61]{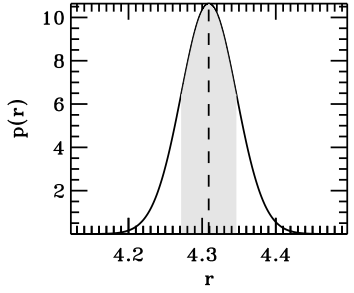}\includegraphics[scale=0.61]{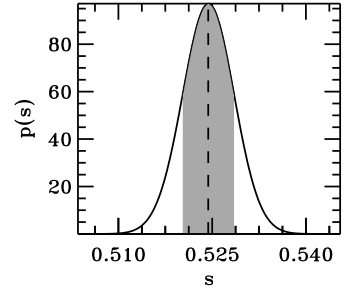}\includegraphics[scale=0.61]{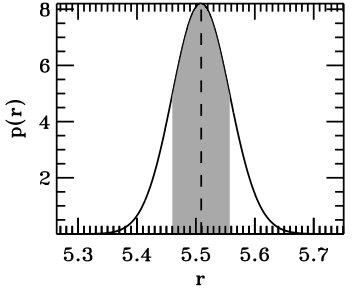}\includegraphics[scale=0.61]{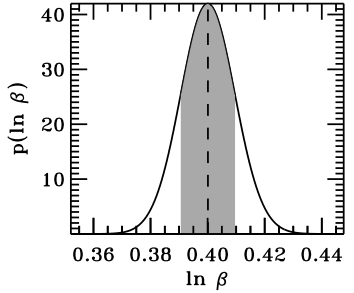}
\caption{Marginal PDFs for the free parameters of the models $\mathcal{M}_1$ and $\mathcal{M}_{1,\beta}$, where expectation values listed in Table~\ref{tab:model} (dashed lines) and 68.3\,\% Bayesian credible regions (light gray for $\mathcal{M}_1$ and dark gray for $\mathcal{M}_{1,\beta}$) have been marked. }
\label{fig:marg_m1}
\end{center}
\end{figure*}

The natural logarithm of model $\mathcal{M}_5$ lead us to
\begin{equation}
\begin{split}
\ln \left( \frac{A^{(4)}_\mathrm{bol}}{\abolsun} \right) =& -2.5 \ln \left( \frac{\nmax}{\nmaxsun} \right) \\
& + 2 \ln \left( \frac{\Dnu}{\Dnusun} \right) + 0.5 \ln \left( \frac{\tauosc}{\tauoscsun} \right) \\
&+ (2.3 -r) \ln \left( \frac{\teff}{\teffsun} \right) \, .
\label{eq:lnmodel5}
\end{split}
\end{equation}
Thus, the total uncertainty for the $i$-th star of the sample is given by
\begin{equation}
\begin{split}
\widetilde{\sigma}^2_i (r) =&\, \widetilde{\sigma}^2_{A_i} + 6.25 \widetilde{\sigma}^2_{\nu_{\mathrm{max},i}} + 4 \widetilde{\sigma}^2_{\Dnu_i} \\
&+ 0.25 \widetilde{\sigma}^2_{\tau_{\mathrm{osc},i}} + (2.3-r)^2 \widetilde{\sigma}^2_{T_{\mathrm{eff},i}} \, ,
\label{eq:tot_uncert_model5}
\end{split}
\end{equation}
with $\widetilde{\sigma}^2_{\tau_{\mathrm{osc},i}} \equiv \sigma_{\tau_{\mathrm{osc},i}} / \tau_{\mathrm{osc},i}$, and is the same for both the models here considered, as model $\mathcal{M}_{5,\beta}$ differs only by the additional term $\ln \beta$. The resulting models are shown in Figure~\ref{fig:model5}, with similar descriptions as those adopted for Figures~\ref{fig:marg_m1} and~\ref{fig:model1}.

\subsection{Models $\mathcal{M}_6$ and $\mathcal{M}_{6,\beta}$}
Following the same arguments used for the other models, we obtain
\begin{equation}
\begin{split}
\ln \left( \frac{A^{(5)}_\mathrm{bol}}{\abolsun} \right) = & \, \, (2-3t) \ln \left( \frac{\nmax}{\nmaxsun} \right) \\
&+ (4t -4) \ln \left( \frac{\Dnu}{\Dnusun} \right) + 0.5 \ln \left( \frac{\tauosc}{\tauoscsun} \right)\\
&+ (4.55 -r -1.5t) \ln \left( \frac{\teff}{\teffsun} \right) \, .
\label{eq:lnmodel6}
\end{split}
\end{equation}
for the model $\mathcal{M}_6$. The new uncertainties to be considered will depend on the free parameters $r$ and $t$, thus we have
\begin{equation}
\begin{split}
\widetilde{\sigma}^2_i (r, t) =&\, \widetilde{\sigma}^2_{A_i} + (2 -3t)^2 \widetilde{\sigma}^2_{\nu_{\mathrm{max},i}} + (4t - 4)^2 \widetilde{\sigma}^2_{\Dnu_i} \\
&+ 0.25 \widetilde{\sigma}^2_{\tau_{\mathrm{osc},i}} + (4.55 -r - 1.5t)^2 \widetilde{\sigma}^2_{T_{\mathrm{eff},i}} \, .
\label{eq:tot_uncert_model6}
\end{split}
\end{equation}
 with the same quantities adopted in Eq.~(\ref{eq:tot_uncert_model5}). An analogous discussion to that used for other scaling relations has to be applied for model $\mathcal{M}_{6,\beta}$. The results of the inference for both models are plotted in Figure~\ref{fig:model6}.
 
\section{Model comparison}
\label{sec:evidence}
As mentioned in Section~\ref{sec:inference}, the term $p(A \mid \mathcal{M})$ appearing in Eq.~(\ref{eq:bayes}) (Bayesian evidence) is the one of interest for solving the problem of model comparison in the context of Bayesian statistics \citep[e.\,g. see][]{Trotta08,Benomar09,Handberg11,Gruber12}. The Bayesian evidence is given by integrating the numerator appearing in the right-hand side of Eq.~(\ref{eq:bayes}) over all the possible values of the free parameters $\xi_1, \xi_2, \dots, \xi_k$. Thus we have
\begin{equation}
\mathcal{E}_\mathcal{M} \equiv p ( A \mid \mathcal{M} ) = \int_{\Omega_\mathcal{M}} \mathcal{L} (\boldsymbol{\xi}) \pi ( \boldsymbol{\xi} \mid \mathcal{M} ) d \boldsymbol{\xi} \, ,
\label{eq:evidence}
\end{equation}
where $\Omega_{\mathcal{M}}$ represents the parameter space, defined by the intervals of variation of the free parameters that formalize the hypotheses of the model $\mathcal{M}$, and having volume given by Eq.~(\ref{eq:prior_tot}).  
The Bayesian evidence given by Eq.~(\ref{eq:evidence}) basically represents the integral of the likelihood function ÒaveragedÓ by the prior distribution. As the prior $\pi(\boldsymbol{\xi} \mid \mathcal{M})$ has to be normalized, the evidence depends on the parameter space. Thus, the intervals $[\xi_j^{\rm min}, \xi_j^{\rm max}]$ of the free parameters $\xi_j$ used for computing Eq.~(\ref{eq:evidence}) are those listed in Table~\ref{tab:intervals_amp}.

Since a measure of $\mathcal{E}_\mathcal{M}$ alone does not carry any meaningful information, to solve the problem of model comparison it is useful to take into account the ratios (or odds) of the evidences, namely the so-called Bayes factor, which is given as
\begin{equation}
B_{ij} = \frac{p (A \mid \mathcal{M}_i)}{p (A \mid \mathcal{M}_j)} = \frac{\mathcal{E}_{\mathcal{M}_i}}{\mathcal{E}_{\mathcal{M}_j}} \, .
\label{eq:bayes_factor_amp}
\end{equation}
In case $B_{ij} > 1$ the model $\mathcal{M}_i$ is the favored one, while conversely if $B_{ij} < 1$ the model $\mathcal{M}_j$ ought to be preferred. The resulting natural logarithms of the Bayes factor, which are computed according to Eq.~(\ref{eq:bayes_factor_amp}) are listed in Tables~\ref{tab:bayes},~\ref{tab:bayessc}, and~\ref{tab:bayeslc}, for the cases of the entire sample, and SC and LC targets, respectively. Therefore, if $\ln B_{ij} > 0$ the model $M_i$ is preferred over $M_j$ and vice versa if $\ln B_{ij} < 0$.

\begin{table*}
\begin{minipage}{160mm}
\centering
 \caption{Natural logarithms of the Bayes factor $\ln B_{ij}$ for each pair of models $\mathcal{M}_i$, $\mathcal{M}_j$ as derived by means of Eq.~ (\ref{eq:bayes_factor_amp}) for the case of the entire sample.}
\begin{tabular}{lrrrrrrrr}
  \hline	
  \\[-8pt]
Model & \multicolumn{1}{c}{$\mathcal{M}_1$} & \multicolumn{1}{c}{$\mathcal{M}_{1,\beta}$} & \multicolumn{1}{c}{$\mathcal{M}_2$} & \multicolumn{1}{c}{$\mathcal{M}_4$} & \multicolumn{1}{c}{$\mathcal{M}_{4,\beta}$} & \multicolumn{1}{c}{$\mathcal{M}_5$} & \multicolumn{1}{c}{$\mathcal{M}_{5,\beta}$} & \multicolumn{1}{c}{$\mathcal{M}_6$}\\[1pt]
  \hline
  \\[-8pt]
  $\mathcal{M}_1$ &      \multicolumn{1}{c}{--}  & & & & & & & \\
  $\mathcal{M}_\mathrm{1,\beta}$ & $1038.2$  &   \multicolumn{1}{c}{--}   & & & & & &\\
  $\mathcal{M}_2$ &     $-626.7$  & $-1664.9$   & \multicolumn{1}{c}{--}  & & & & &\\
  $\mathcal{M}_4$ &    $2582.3$   &  $1544.1$   &  $3209.0$    &  \multicolumn{1}{c}{--}  & & & &\\
\cellcolor[gray]{0.9}$\mathcal{M}_\mathrm{4,\beta}$ & \cellcolor[gray]{0.9}$3690.6$   &  \cellcolor[gray]{0.9}$2652.4$   &  \cellcolor[gray]{0.9}$4317.3$   &  \cellcolor[gray]{0.9}$1108.3$   &   \multicolumn{1}{c}{--} & & &\\
  $\mathcal{M}_5$ &     $2613.9$    & $1575.7$   &  $3240.6$    &   $31.6$   &  \cellcolor[gray]{0.9}$-1076.6$   &   \multicolumn{1}{c}{--}   & & \\
  $\mathcal{M}_\mathrm{5,\beta}$ &     $2748.4$   &  $1710.2$   &  $3375.2$   &   $166.2$  &  \cellcolor[gray]{0.9}$-942.1$   &  $134.5$    &  \multicolumn{1}{c}{--}    & \\
    $\mathcal{M}_6$ &     $2809.1$   &  $1770.9$   &  $3435.9$    &  $226.8$   & \cellcolor[gray]{0.9}$-881.4$    &  $195.2$   &   $60.7$   &   \multicolumn{1}{c}{--} \\
  $\mathcal{M}_\mathrm{6,\beta}$ &     $2810.0$   &  $1771.8$   &  $3436.7$   &  $227.7$   &  \cellcolor[gray]{0.9}$-880.6$    &  $196.1$    &   $61.6$    &   $0.9$\\[2pt]  
  \hline
 \end{tabular}
\label{tab:bayes}
\end{minipage}
\end{table*}
\begin{table*}
\begin{minipage}{160mm}
\centering
 \caption{Same description as for Table~\ref{tab:bayes} but in the case of the sample of SC targets.}
\begin{tabular}{lrrrrrrrrr}
  \hline	
  \\[-8pt]
Model & \multicolumn{1}{c}{$\mathcal{M}_1$} & \multicolumn{1}{c}{$\mathcal{M}_{1,\beta}$} & \multicolumn{1}{c}{$\mathcal{M}_2$} & \multicolumn{1}{c}{$\mathcal{M}_4$} & \multicolumn{1}{c}{$\mathcal{M}_{4,\beta}$} & \multicolumn{1}{c}{$\mathcal{M}_5$} & \multicolumn{1}{c}{$\mathcal{M}_{5,\beta}$} & \multicolumn{1}{c}{$\mathcal{M}_6$}\\[1pt]
  \hline
  \\[-8pt]
  $\mathcal{M}_1$ &  \multicolumn{1}{c}{--}  & & & & & & & \\
  $\mathcal{M}_\mathrm{1,\beta}$ &      $118.8$    &   \multicolumn{1}{c}{--} & & & & & & \\
  $\mathcal{M}_2$ &     $-122.4$   &  $-241.2$    &    \multicolumn{1}{c}{--}   & & & & & \\
  $\mathcal{M}_4$ &      $225.2$   &  $106.4$   &   $347.6$    &    \multicolumn{1}{c}{--}   & & & & \\
\cellcolor[gray]{0.9}$\mathcal{M}_\mathrm{4,\beta}$ & \cellcolor[gray]{0.9}$334.7$   &   \cellcolor[gray]{0.9}$215.8$   &   \cellcolor[gray]{0.9}$457.0$   &   \cellcolor[gray]{0.9}$109.5$    &    \multicolumn{1}{c}{--}   & & & \\
  $\mathcal{M}_5$ &      $243.7$   &   $124.8$   &   $366.0$    &   $18.5$   &   \cellcolor[gray]{0.9}$-91.0$   &     \multicolumn{1}{c}{--}  & &\\
  $\mathcal{M}_\mathrm{5,\beta}$ &      $242.7$   &  $123.9$   &   $365.1$    &   $17.5$   &   \cellcolor[gray]{0.9}$-91.9$   &    $-0.9$    &    \multicolumn{1}{c}{--}    & \\
    $\mathcal{M}_6$ &      $244.3$  &  $125.5$   &   $366.7$    &   $19.1$   &   \cellcolor[gray]{0.9}$-90.3$    &    $0.7$    &    $1.6$   &     \multicolumn{1}{c}{--}  \\
  $\mathcal{M}_\mathrm{6,\beta}$ &      $261.4$  &  $142.6$   &   $383.8$    &   $36.2$   &   \cellcolor[gray]{0.9}$-73.2$    &   $17.8$   &    $18.7$    &   $17.1$  \\[2pt]
  \hline
 \end{tabular}
\label{tab:bayessc}
\end{minipage}
\end{table*}
\begin{table*}
\begin{minipage}{160mm}
\centering
 \caption{Same description as for Table~\ref{tab:bayes} but in the case of the sample of LC targets.}
\begin{tabular}{lrrrrrrrr}
  \hline	
  \\[-8pt]
Model & \multicolumn{1}{c}{$\mathcal{M}_1$} & \multicolumn{1}{c}{$\mathcal{M}_{1,\beta}$} & \multicolumn{1}{c}{$\mathcal{M}_2$} & \multicolumn{1}{c}{$\mathcal{M}_4$} & \multicolumn{1}{c}{$\mathcal{M}_{4,\beta}$} & \multicolumn{1}{c}{$\mathcal{M}_5$} & \multicolumn{1}{c}{$\mathcal{M}_{5,\beta}$} & \multicolumn{1}{c}{$\mathcal{M}_6$}\\[1pt]
  \hline
  \\[-8pt]
  $\mathcal{M}_1$ &        \multicolumn{1}{c}{--}  & & & & & & & \\
  $\mathcal{M}_\mathrm{1,\beta}$ &       $59.2$    &    \multicolumn{1}{c}{--}  & & & & & & \\
  $\mathcal{M}_2$ &       $30.2$  &    $-29.0$   &     \multicolumn{1}{c}{--}  & & & & & \\
  $\mathcal{M}_4$ &    $1003.2$   &   $943.9$   &   $973.0$    &    \multicolumn{1}{c}{--}   & & & & \\
\cellcolor[gray]{0.9}$\mathcal{M}_\mathrm{4,\beta}$ &  \cellcolor[gray]{0.9}   $1094.2$   & \cellcolor[gray]{0.9} $1035.0$   & \cellcolor[gray]{0.9} $1064.0$   &  \cellcolor[gray]{0.9}  $91.1$   &     \multicolumn{1}{c}{--} & & & \\
  $\mathcal{M}_5$ &     $445.7$   &   $386.4$   &   $415.5$  &   $-557.5$   &  \cellcolor[gray]{0.9}$-648.6$     &   \multicolumn{1}{c}{--} & & \\
  $\mathcal{M}_\mathrm{5,\beta}$ &      $491.5$   &   $432.3$   &   $461.3$  &   $-511.6$   &  \cellcolor[gray]{0.9}$-602.7$    &   $45.9$   &     \multicolumn{1}{c}{--}   & \\
    $\mathcal{M}_6$ &      $444.9$    &  $385.6$   &   $414.7$  &   $-558.3$   &  \cellcolor[gray]{0.9}$-649.4$    &   $-0.8$   &   $-46.7$   &     \multicolumn{1}{c}{--} \\
  $\mathcal{M}_\mathrm{6,\beta}$ &      $490.9$   &   $431.6$    &  $460.7$   &  $-512.3$   &  \cellcolor[gray]{0.9}$-603.4$    &   $45.2$    &   $-0.7$    &   $46.0$  \\[2pt]
  \hline
 \end{tabular}
\label{tab:bayeslc}
\end{minipage}
\end{table*}
It is sometimes useful to consider so-called Information Criteria, which may offer a simpler alternative to the Bayesian evidence, whose numerical computation in some cases can be very time demanding. In particular, we adopted the Bayesian Information Criterion (BIC), also known as Schwarz Information Criterion \citep{schwarz}, which follows from a Gaussian approximation to the Bayesian evidence in the limit of a large sample size, as it can be represented by our sample of stars ($N \gg 1$). Thus, the BIC reads
\begin{equation}
\mbox{BIC} \equiv -2 \Lambda_\mathrm{max} + k \ln N \, ,
\label{eq:amp_BIC}
\end{equation}
where $k$ is the number of free parameters of the model considered (i.\,e. the dimension of the corresponding parameter space), and $N$ the number of data points. Since $\Lambda_\mathrm{max}$ is known, the BIC can be computed straightforwardly. The resulting values of the BIC are listed in Table~\ref{tab:bic} for the cases of the entire sample (second column) and of SC and LC targets separately (third and fourth columns, respectively). According to the Occam's razor principle on which Bayesian model comparison relies, the most eligible model is the one that minimizes the BIC. 

As highlighted by the shaded rows and columns, the model $\mathcal{M}_{4,\beta}$ is largely the favored one for all the samples considered because its evidence is always greater than those of any other model investigated in this work. In addition, the BIC confirms the
result computed through the evidences. 

\begin{table}
\centering
 \caption{Bayesian Information Criterion (BIC) computed for all the models in three cases considered: all targets (second column), SC targets (third column), LC targets (fourth column).}
\begin{tabular}{lrrr}
  \hline
  \\[-8pt]
  Model & \multicolumn{1}{c}{BIC} & \multicolumn{1}{c}{BIC$^\mathrm{(SC)}$}& \multicolumn{1}{c}{BIC$^\mathrm{(LC)}$}\\[1pt]
  \hline
  \\[-8pt]
  $\mathcal{M}_1$ & 7081 & 655 & 1612 \\
  $\mathcal{M}_\mathrm{1,\beta}$ & 5004 & 417 & 1497 \\
  $\mathcal{M}_2$ & 8335 & 901 & 1552 \\
  $\mathcal{M}_4$ & 1920 & 209 & $-393$ \\
\rowcolor[gray]{0.9}  $\mathcal{M}_\mathrm{4,\beta}$ & $-296$ & $-11$ & $-574$ \\
  $\mathcal{M}_5$ & 1859 & 172 & 723 \\
  $\mathcal{M}_\mathrm{5,\beta}$ & 1589 & 173 & 632 \\
    $\mathcal{M}_6$ & 1469 & 173 & 726 \\
  $\mathcal{M}_\mathrm{6,\beta}$ & 1468 & 137 & 635 \\[2pt]
  \hline
 \end{tabular}
\label{tab:bic}
\end{table}

\section{Discussion}
\label{sec:amp_discussion}
The analysis described in Section~\ref{sec:inference} and in Section~\ref{sec:evidence} lead us to interesting results about the use of the amplitude scaling relations in asteroseismology. The main aspects of the work presented here can be divided into two groups, whose contents we discuss in the following.

\subsection{Results from Bayesian parameter estimation}
\label{sec:discuss_inference}
\begin{figure*}
\begin{center}
\includegraphics[scale=0.45]{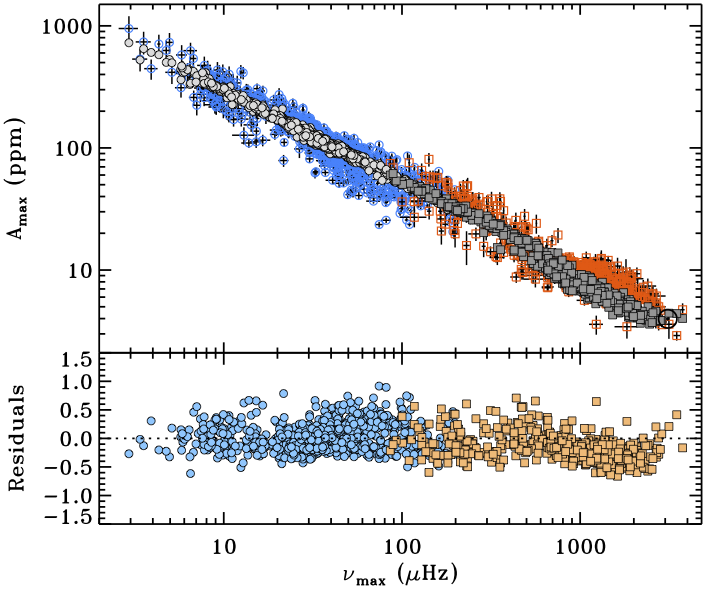}\includegraphics[scale=0.45]{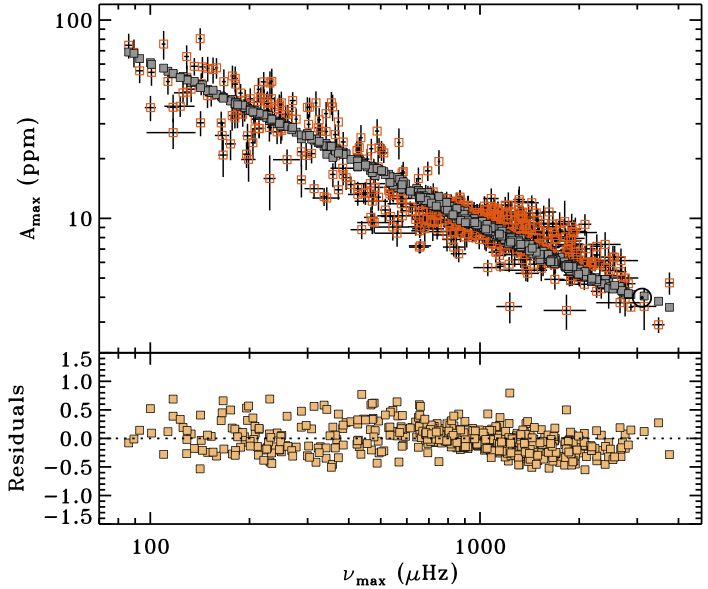}\includegraphics[scale=0.45]{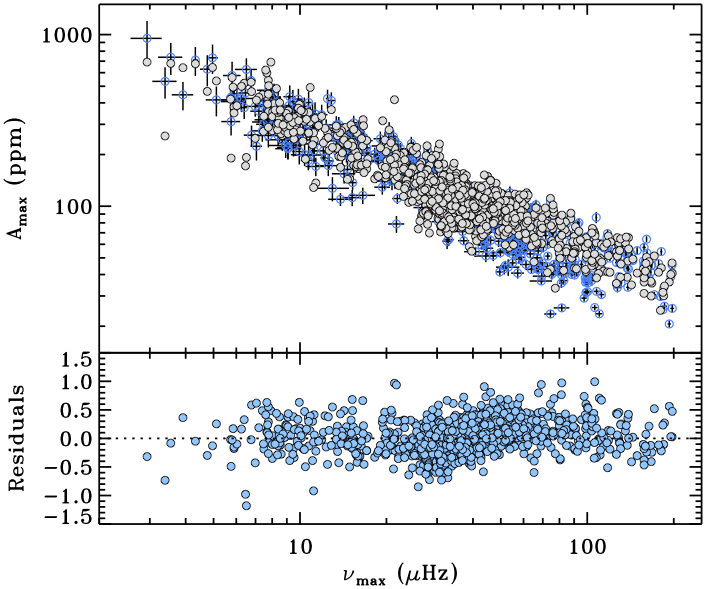}
\includegraphics[scale=0.45]{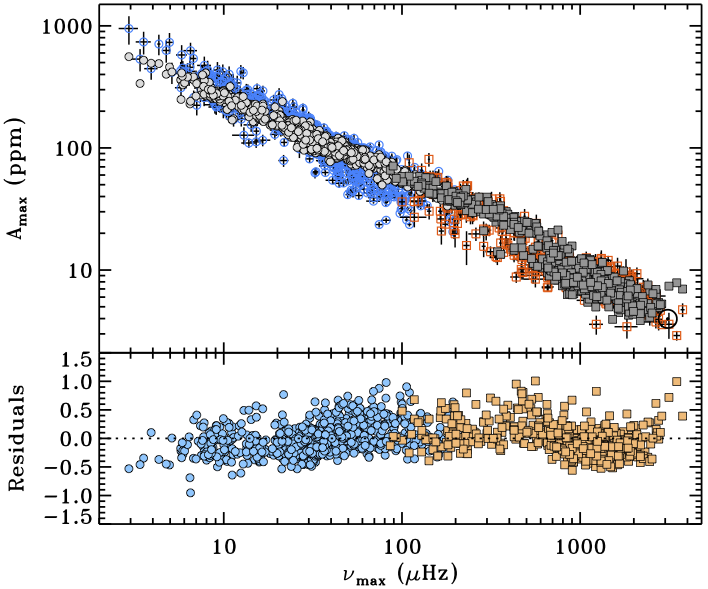}\includegraphics[scale=0.45]{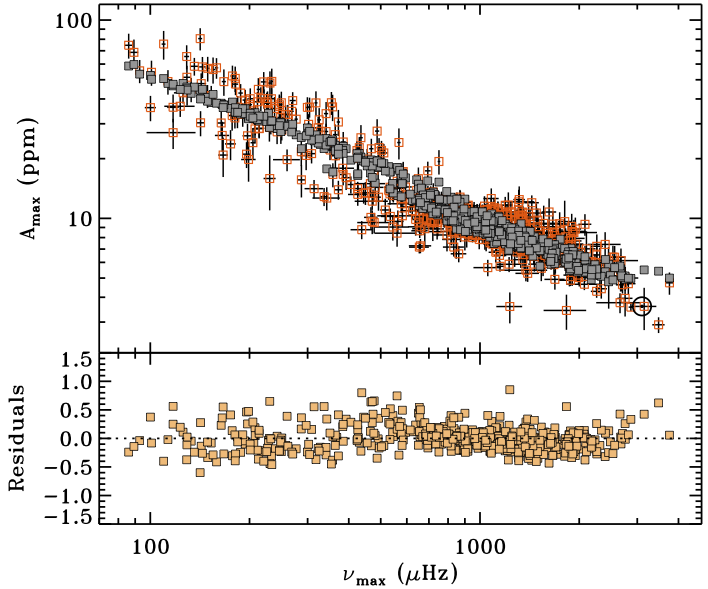}\includegraphics[scale=0.45]{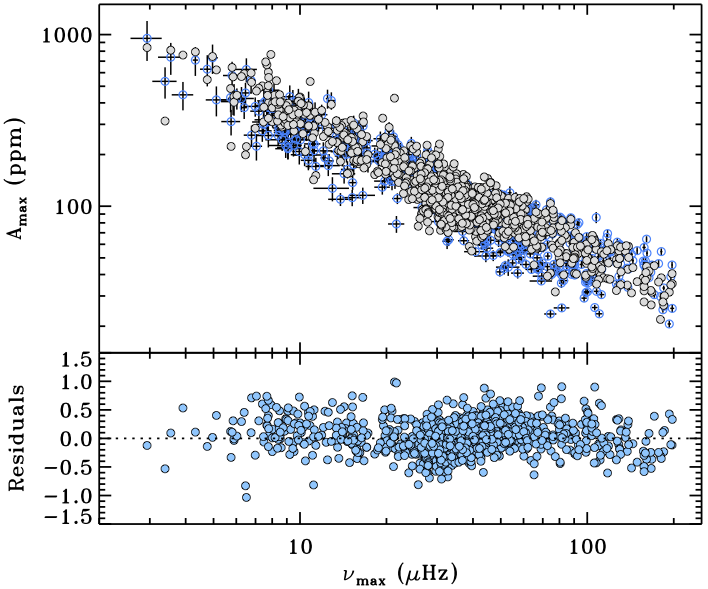}
\caption{\textit{Top panels}: Predicted amplitudes (filled light-gray circles for LC targets and filled dark-gray squares for SC targets) for model $\mathcal{M}_1$ plotted against $\nmax$ in the three cases considered (all sample in left panel, SC targets only in middle panel, LC targets only in right panel). The expectation values of the free parameters reported in Tables~\ref{tab:model},~\ref{tab:modelsc},~\ref{tab:modellc}, have been adopted for plotting the predicted amplitudes. Observed amplitudes are shown in background for both SC (open orange squares) and LC (open blue circles) targets, together with 1-$\sigma$ error bars shown on both quantities. The Sun's symbol ($\sun$) is added for comparison, where $A_{650,\odot} = 3.98\,$ppm ($\lambda = 650\,$nm). The residuals arising from the difference between the logarithms of observed and predicted amplitudes, according to Eqs.~(\ref{eq:lnmodel1}) and~(\ref{eq:lnmodel1off}), are also plotted (same symbols). \textit{Bottom panels}: Same description of the top panels but for the model $\mathcal{M}_{1,\beta}$.}
\label{fig:model1}
\end{center}
\end{figure*}

\begin{figure*}
\begin{center}
\includegraphics[scale=0.46]{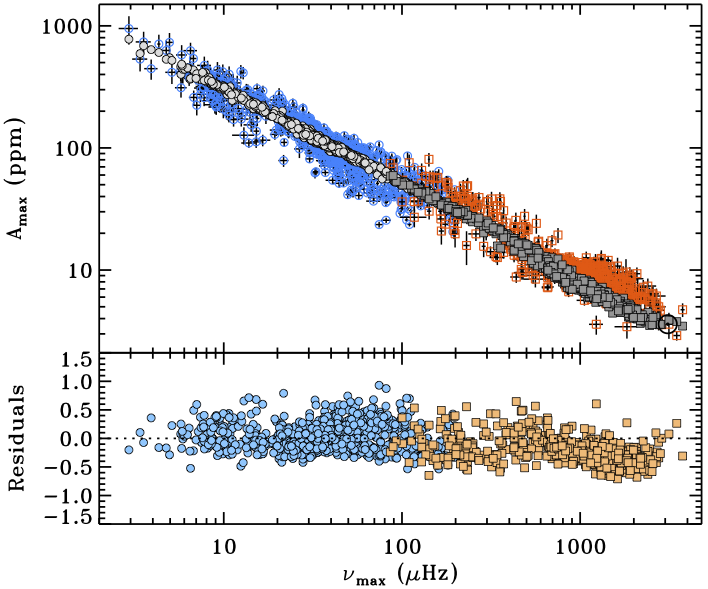}\includegraphics[scale=0.46]{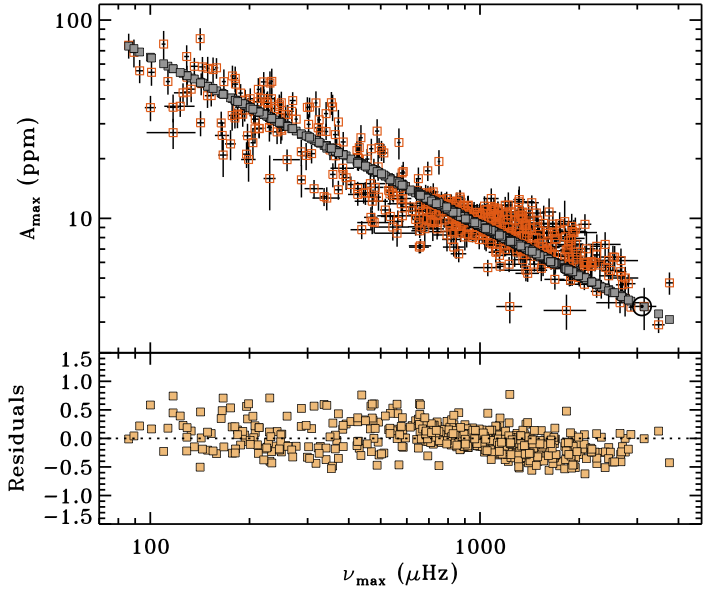}\includegraphics[scale=0.46]{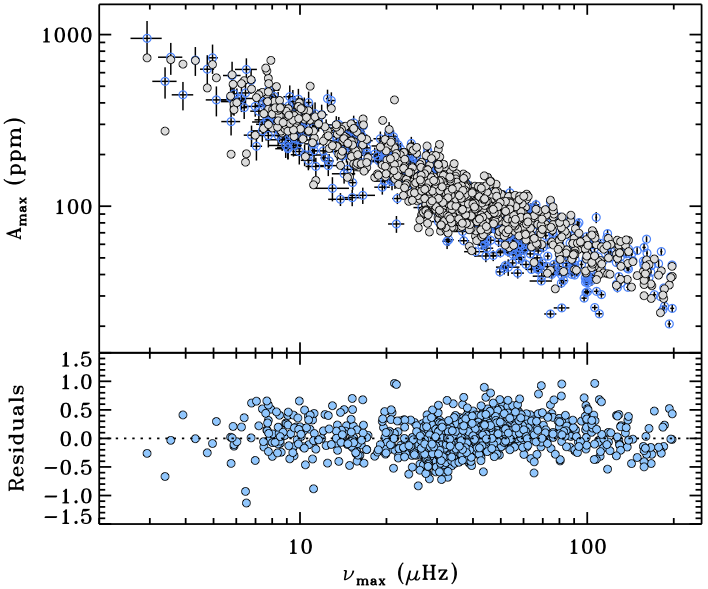}
\caption{Same description as for Figure~\ref{fig:model1} but for the model $\mathcal{M}_2$.}
\label{fig:model2}
\end{center}
\end{figure*}

\begin{figure*}
\begin{center}
\includegraphics[scale=0.46]{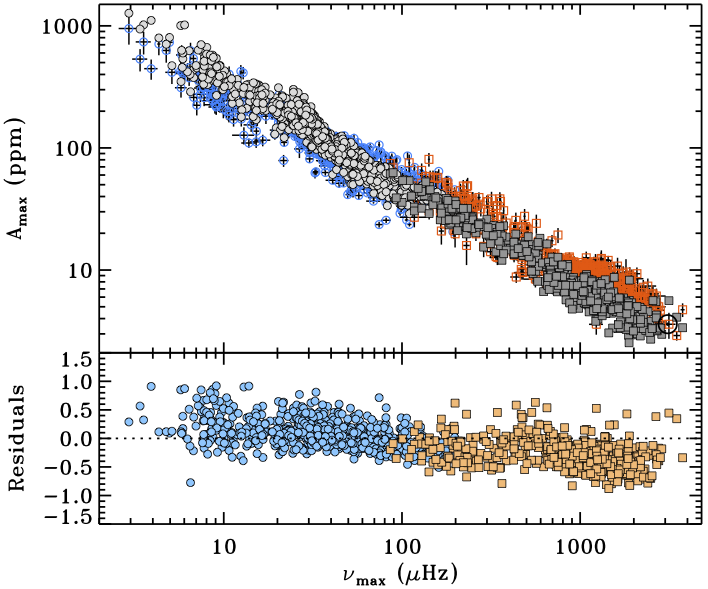}\includegraphics[scale=0.46]{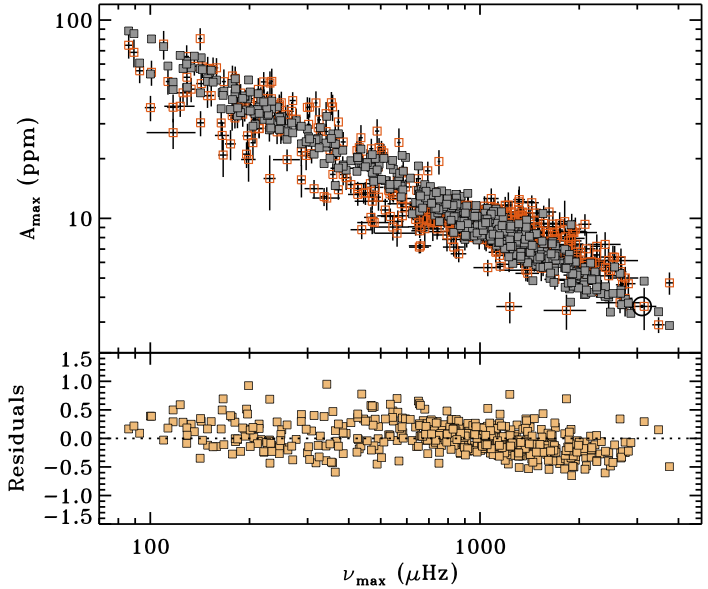}\includegraphics[scale=0.46]{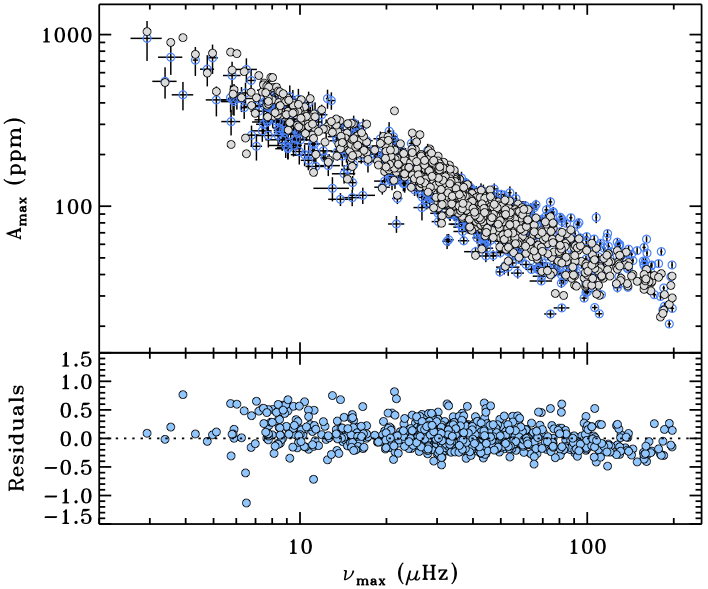}
\includegraphics[scale=0.46]{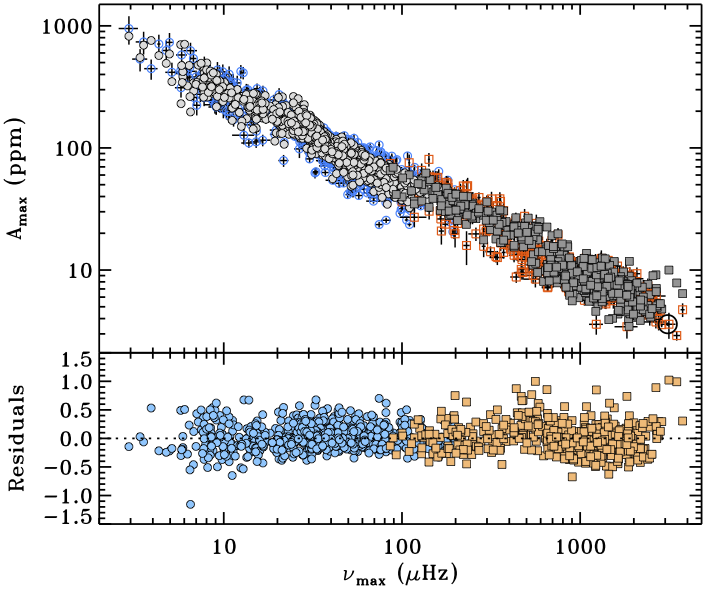}\includegraphics[scale=0.46]{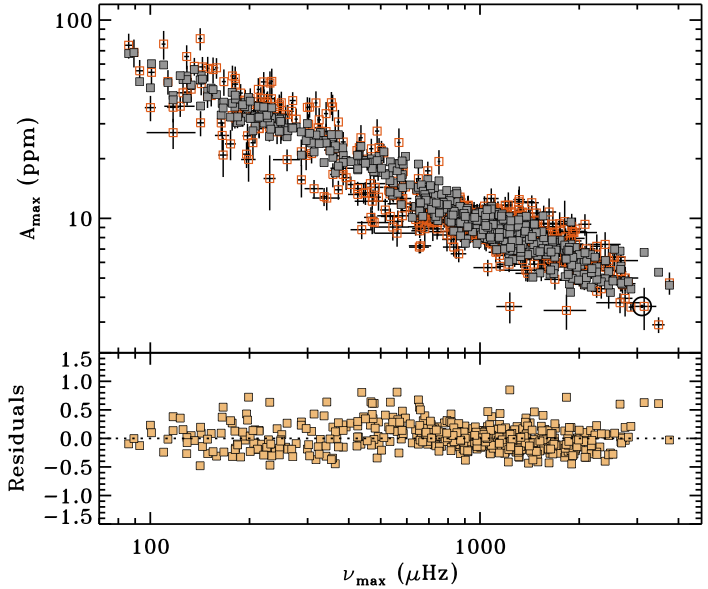}\includegraphics[scale=0.46]{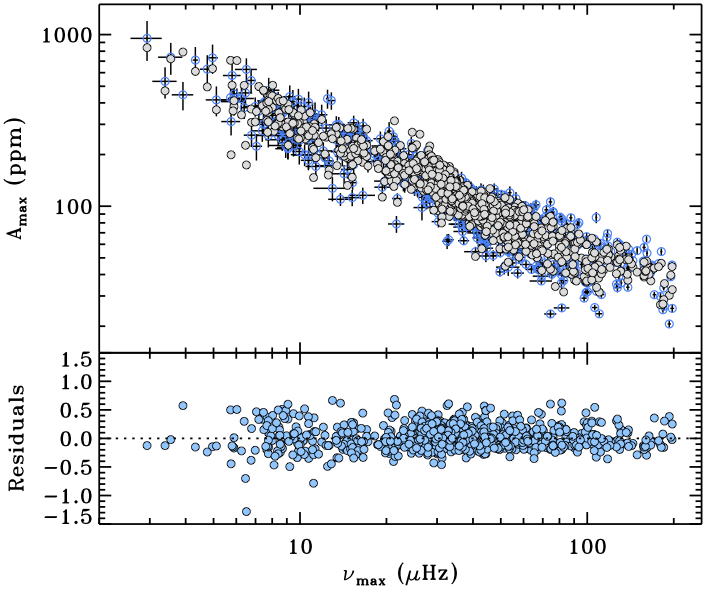}
\caption{Same description as for Figure~\ref{fig:model1} but for the models $\mathcal{M}_4$ (top panels) and $\mathcal{M}_{4,\beta}$ (bottom panels).}
\label{fig:model4}
\end{center}
\end{figure*}

\begin{figure*}
\begin{center}
\includegraphics[scale=0.46]{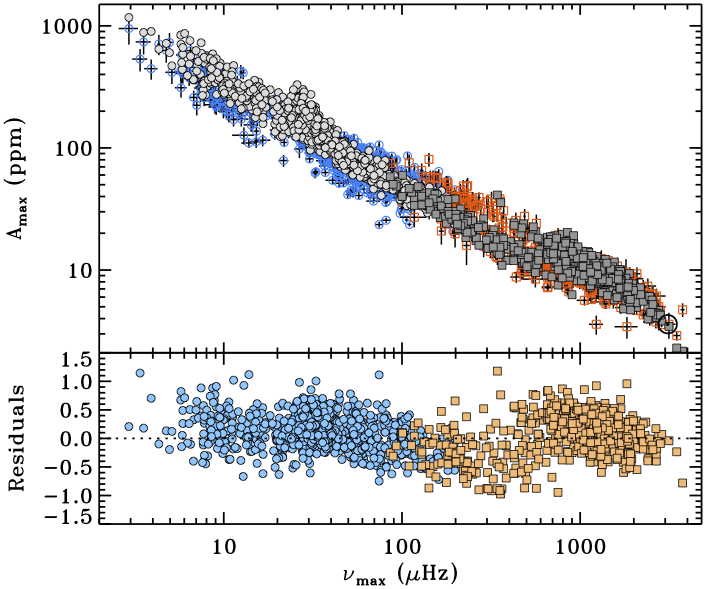}\includegraphics[scale=0.46]{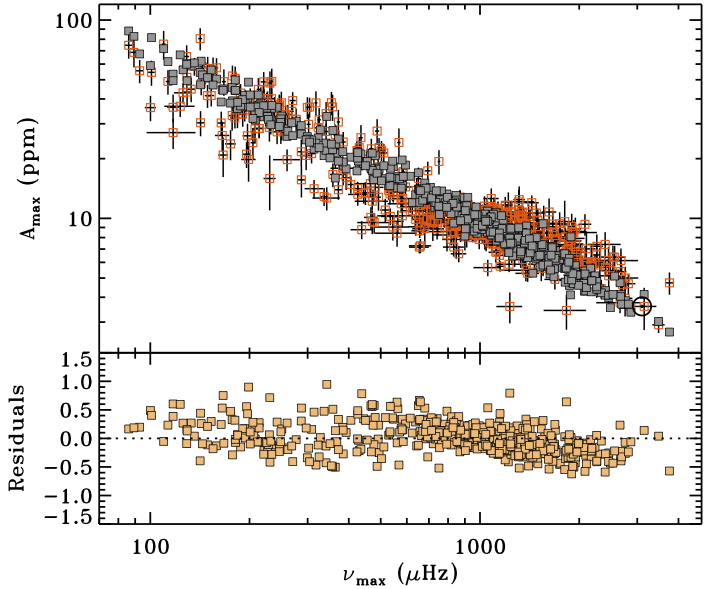}\includegraphics[scale=0.46]{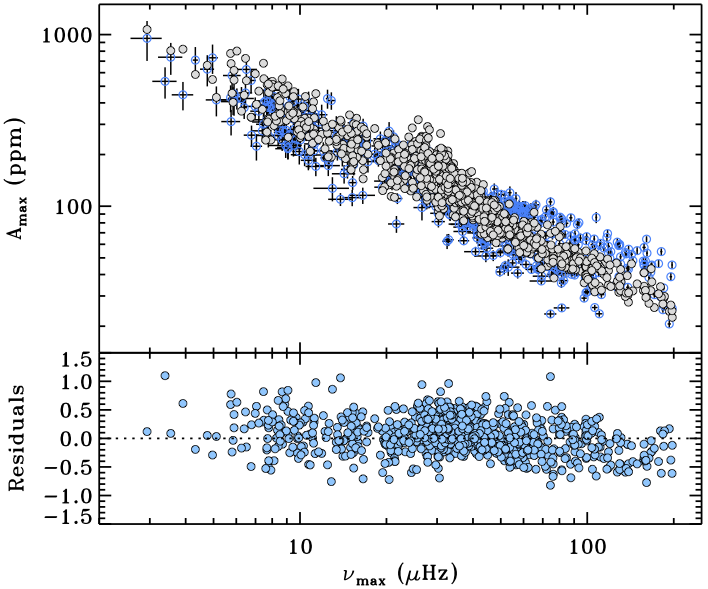}
\includegraphics[scale=0.46]{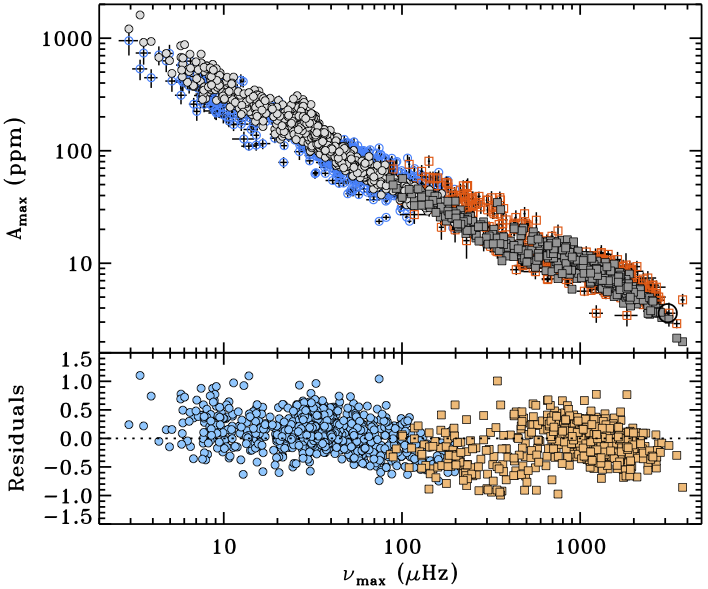}\includegraphics[scale=0.46]{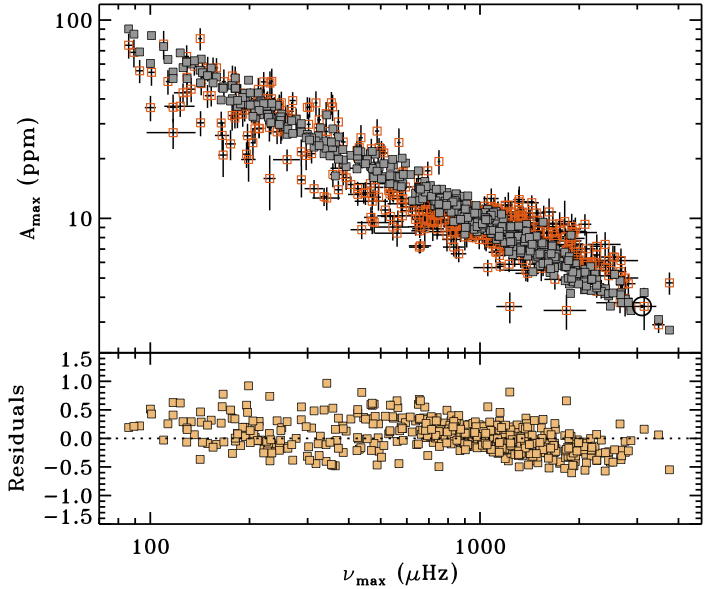}\includegraphics[scale=0.46]{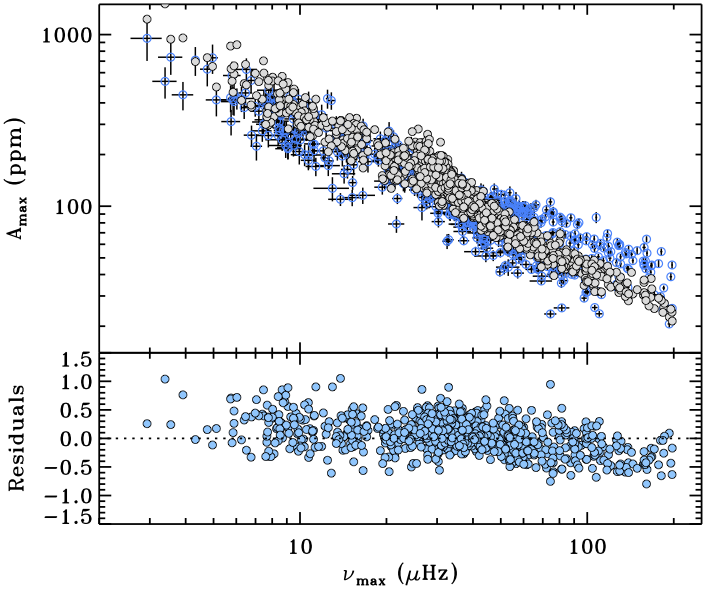}
\caption{Same description as for Figure~\ref{fig:model1} but for the models $\mathcal{M}_5$ (top panels) and $\mathcal{M}_{5,\beta}$ (bottom panels).}
\label{fig:model5}
\end{center}
\end{figure*}

\begin{figure*}
\begin{center}
\includegraphics[scale=0.46]{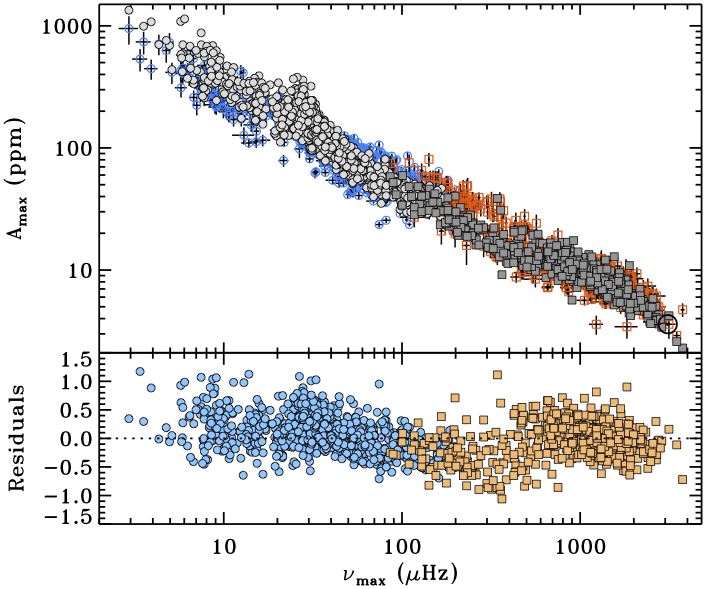}\includegraphics[scale=0.46]{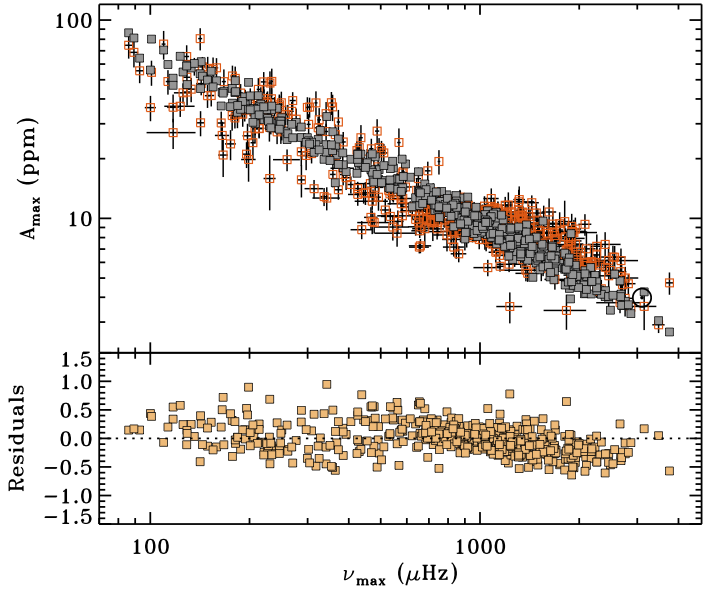}\includegraphics[scale=0.46]{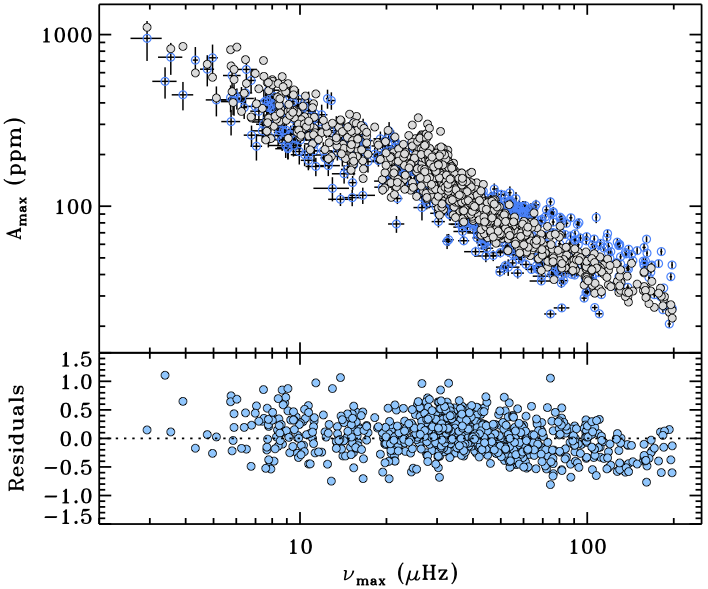}
\includegraphics[scale=0.46]{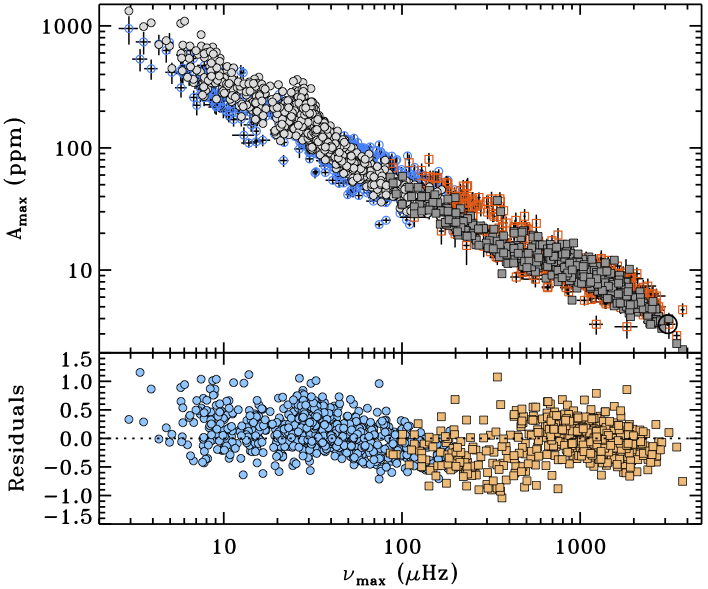}\includegraphics[scale=0.46]{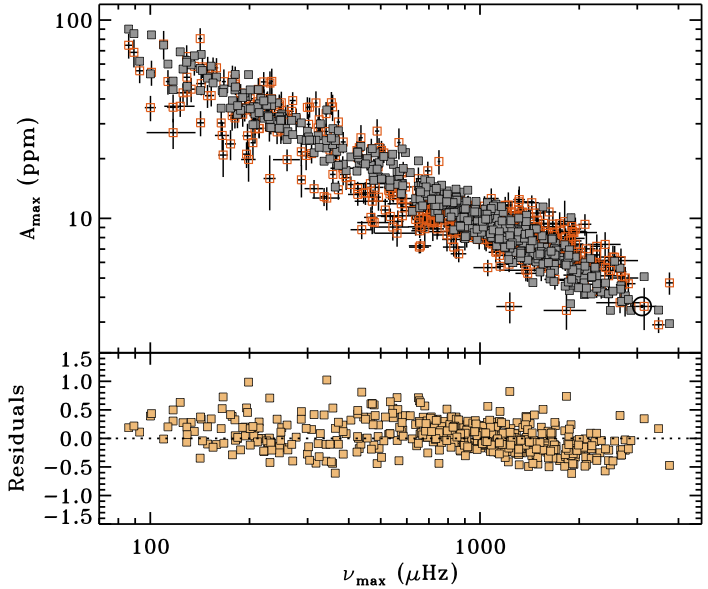}\includegraphics[scale=0.46]{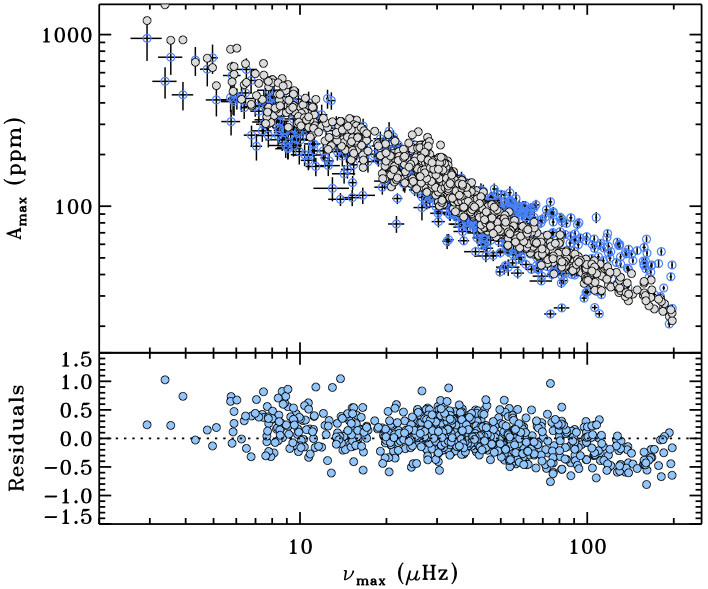}
\caption{Same description as for Figure~\ref{fig:model1} but for the models $\mathcal{M}_6$ (top panels) and $\mathcal{M}_{6,\beta}$ (bottom panels).}
\label{fig:model6}
\end{center}
\end{figure*}

The results coming from the inference described in Section~\ref{sec:inference} show that:
\begin{enumerate}
\item for models from $\mathcal{M}_1$ to $\mathcal{M}_{4,\beta}$, the expectation values of exponent $r$, which we remind was introduced for converting radial velocity amplitudes into photometric ones, differ from the value $r = 2$ adopted for non-adiabatic oscillations \citep{KB95,Stello11,Samadi12}. On one hand, this outcome is even more evident in the case of RGs (namely the LC targets), where we found $r \simeq 10.5$ for model $\mathcal{M}_2$ (Table~\ref{tab:modellc}). Although such a high value for $r$ could be partly explained by a very tight anti-correlation with the exponent $s$ (see Tables~\ref{tab:correlation} and~\ref{tab:correlationlc}), this seems to support the recent findings by \cite{Samadi12} who suggest that for RGs the non-adiabatic effects become significantly important in the driving mechanism of solar-like oscillations and that, in general, the excitation model is underestimating the true amplitudes. On the other hand, in the case of SC targets, which are essentially dominated by MS stars, the mean $r$ is not as correlated with $s$ as for the LC targets (see Table~\ref{tab:correlationsc}), and its estimated values are considerably lower ($r \approx 3$), and much closer to the value $r = 2$ adopted in previous works (Table~\ref{tab:modelsc}). This again could suggest that solar-like oscillations are more adiabatic in early stages of stellar evolution.

Conversely, $r$ becomes negative for the models that take into account the granulation power and the lifetime of the modes ($\mathcal{M}_5$ to $\mathcal{M}_{6,\beta}$), which give a different contribution to the predicted amplitudes from effective temperature (see Sections~\ref{sec:amp_m5} and~\ref{sec:amp_m6}). The reason our estimates of $r$ are much lower than the theoretical value is mainly related to the fact that predicted amplitudes based on the scaling relation by \cite{KB11} are considerably greater than the observed ones, even for the most adiabatic case corresponding to $r = 1.5$ \citep[e.\,g. see the discussions by][]{Stello11,Huber11}. This overestimation could be explained by a missing term containing the mode masses, as proved by recent theoretical calculations by \cite{Samadi12} for a sample of RGs observed by \textit{CoRoT}.

Correlations among the free parameters are very tight in many cases and can be partially responsible of decreasing $r$, except for the case of SC targets, where no significant correlations are found

\item for the models that include the exponent $s$ ($\mathcal{M}_1$ to $\mathcal{M}_{4,\beta}$), the expectation values derived here are fairly compatible with previous results \citep[e.\,g. see][]{KB95,Gilliland08,Dz10,Stello10,Verner11,Baudin11,Stello11,Huber11}. However, it is worth mentioning that when moving from the main sequence to the red giant phase, $s$ decreases, which is apparent for all the models mentioned above. This effect can be partially explained by a compensation of the exponent $r$

\item the expectation values of the exponent $t$ found for models $\mathcal{M}_4, \mathcal{M}_6$, and $\mathcal{M}_{6,\beta}$ are not far from the value $t = 1.7 \pm 0.1$ derived by \cite{Stello11} using a sample of cluster RGs, although correlation effects with the exponents $s$ and $r$ cannot be neglected and are, in some cases, quite pronounced (see Table~\ref{tab:correlation}). For model $\mathcal{M}_{4,\beta}$ instead, we found $t$ to be very close to the value $t = 1.32 \pm 0.02$ derived by \cite{Huber11}, who used a very similar sample of stars but with effective temperatures from KIC. These results are confirmed for both the entire sample and the sample of SC targets. The RGs sample is instead behaving differently, showing values of $t$ significantly lower than the other two samples. This result is certainly dominated by correlations, which are stronger than those of the MS-dominated sample (see below)

\item as a qualitative result we can state that, in general, all the expectations of the free parameters involved in the amplitude scaling relations investigated here are on average rather different when comparing the MS-dominated sample to the RGs one. The correlations among the free parameters are enhancing the differences between dwarfs and giants and are likely to be stronger in the sample of more evolved stars because for red giants there is a large degree of degeneracy in the stellar fundamental properties as stars with different mass all converge along the red giant branch spanning only a small temperature range. As a consequence, this could suggest that not a single of the amplitude scaling relations discussed can be adopted to both MS and RGs simultaneously because the driving and damping mechanisms responsible of generating the observed amplitudes encounter a substantial change as the stars evolve

\item the introduction of a set of models that take into account the offset $\ln \beta$ allowed us to produce new outcomes that provide better fits for the entire set of scaling relations adopted, as it appears clear by looking at the comparison of different fits shown in Figures~\ref{fig:model1},~\ref{fig:model4}, ~\ref{fig:model5}, and ~\ref{fig:model6}, together with the hint provided by our estimates of a weighted rms of the residuals listed in Tables~\ref{tab:model},~\ref{tab:modelsc}, and~\ref{tab:modellc}. In fact, in almost all cases, the offset $\ln \beta$ differs from zero significantly. In addition, we note that our estimates of $\sigma^w_\mathrm{rms}$ are fairly consistent with the total relative uncertainty adopted in Eq.~(\ref{eq:likelihood}) for models $\mathcal{M}_4$, $\mathcal{M}_{4,\beta}$, $\mathcal{M}_6$, and $\mathcal{M}_{6,\beta}$, which are the only ones that have a separate dependence upon the mass of the stars (accounted for in the additional free parameter $t$). Models $\mathcal{M}_1$ to $\mathcal{M}_2$ and models $\mathcal{M}_5$ and $\mathcal{M}_{5,\beta}$ have instead a scatter in the residuals that is from $1.5$ to $2$ times larger than the total relative uncertainties. This suggests that the average relative uncertainty on the observed amplitudes alone is considerably smaller than the intrinsic scatter of the residuals given by Eq.~(\ref{eq:delta}), for any of the models investigated. Thus, we suppose that additional contributions to the total uncertainty are missing for models that do not take into account a separate dependence upon the mass of the stars. It is however important to note that this outcome only holds for the set of measurements and uncertainties adopted in the inference.
\end{enumerate}

\subsection{Results from Bayesian model comparison}
\label{sec:discuss_evidence}
From the model comparison applied to the amplitude scaling relations here investigated, we can say that:
\begin{enumerate}
\item models $\mathcal{M}_1$, $\mathcal{M}_{1,\beta}$ and $\mathcal{M}_2$ are not performing well at predicting the amplitudes along the entire range of stars considered. This is clear from the fact that their corresponding Bayes factors are the lowest among the models considered (see Tables~\ref{tab:bayes}, \ref{tab:bayessc}, \ref{tab:bayeslc}), a result that is confirmed by our computation of the BIC from Eq.~(\ref{eq:amp_BIC}), which reaches its highest value for these models (see Table~\ref{tab:bic} in Section~\ref{sec:evidence}). The larger scatter of the residuals arising from these models (up to a factor of $2$) compared to the total relative uncertainties derived from Eqs.~(\ref{eq:tot_uncert_model1}) and~(\ref{eq:tot_uncert_model2}) are supporting our conclusion from the model selection described in Section~\ref{sec:evidence}

\item models from $\mathcal{M}_4$ to $\mathcal{M}_{6,\beta}$, according to Bayesian principles of model selection, are certainly more favored than the others (Tables~\ref{tab:bayes}, \ref{tab:bayessc}, \ref{tab:bayeslc}, and \ref{tab:bic}). In particular, model $\mathcal{M}_{4,\beta}$, which includes a separate dependence upon the mass of the stars (see Section~\ref{sec:mass_amp}), is strongly dominant over all the other scaling relations investigated, by at least a factor of $\sim \exp(880)$ if one takes into account the Bayes factor. Thus, model $\mathcal{M}_{4,\beta}$ is the best one according to the computation of both the evidences and the BIC for all the samples used. This suggests that the spread observed in the amplitudes is likely to be caused by a spread in mass of the stars. This is also confirmed by the consistency between $\sigma^w_\mathrm{rms}$ and the total relative uncertainties computed from Eq.~(\ref{eq:tot_uncert_model4}).

The models having an evidence weaker than that of $\mathcal{M}_{4,\beta}$, but still stronger than that of models from $\mathcal{M}_1$ to $\mathcal{M}_2$, are the models from $\mathcal{M}_5$ to $\mathcal{M}_{6,\beta}$, which instead include the effects of mode lifetimes and granulation power (see Section~\ref{sec:amp_m5}). These results are again confirmed for all the three samples considered from both the evidences and the BIC
\item all the models that take the proportionality term $\beta$ into account, are preferred to their counterparts without this extra free parameter, with the only exception of models $\mathcal{M}_6$ and $\mathcal{M}_{6,\beta}$ for which the model comparison is inconclusive in the case of the entire sample of stars (Table~\ref{tab:bayes}). These results are confirmed from the computation of both the evidences and the BIC regardless of the evolutionary state of the stars. As a consequence, according to the set of measurements and uncertainties adopted in this work, we can conclude that the Sun is not necessarily a good reference star for this \textit{Kepler} sample, and that only models having $\beta \neq 0$ should be considered. Equivalently, one could consider models that do not include $\beta$ as a free parameter, but having a reference amplitude for scaling given by $\beta \abolsun$ (for models $\mathcal{M}_2$ to $\mathcal{M}_{6,\beta}$) or $\beta A_{\lambda,\odot}$ (for models $\mathcal{M}_1$ and $\mathcal{M}_{1,\beta}$), with $\beta$ given according to the results shown in Tables~\ref{tab:model},~\ref{tab:modelsc}, and~\ref{tab:modellc}. Such result could also be supported by the effect of the stellar activity (occurring mostly for dwarfs), which is responsible of reducing the amplitudes of solar-like oscillations \citep[see the discussion by][]{Chaplin11act,Huber11}. In fact, as noted by \cite{Huber11}, the Sun shows on average a higher activity level than the other stars of the sample.
\end{enumerate}

\section{Conclusions}
\label{sec:amp_conclusion}
A Bayesian approach to test the set of amplitude scaling relations discussed in this work allowed us to draw new interesting conclusions on the free parameters that describe all the models. 

First, as evident from the results derived in Section~\ref{sec:evidence}, our analysis strongly recommends the use of Eq.~(\ref{eq:model4}) for predicting amplitudes of solar-like oscillations, for all the stars spanning from MS to RGs. This result, together with the consistency found between our estimates of the fit quality, $\sigma^w_\mathrm{rms}$, arising from the corresponding models and the total relative uncertainties discussed in Section~\ref{sec:model4}, supports the idea that a mass difference from star to star is among the main effects that produce the observed spread in the oscillation amplitudes \citep[see also][]{Huber10redgiant,Huber11,Stello11}.

Second, one should keep in mind that according to the Bayesian parameter estimation described in Section~\ref{sec:inference} the results arising from the inference change considerably from MS to RGs. This suggests that, in general, particular attention has to be paid when using amplitude scaling relations for samples containing a large range of stars. This behavior is already apparent from our plot of the asteroseismic HR diagram (Figure~\ref{fig:sample}, bottom panel), which shows that, assuming the logarithmic amplitudes change linearly with the logarithm of the temperatures --- as done by all the models considered --- the two samples have on average a quite different slope, which is reflected mainly in a different value of the exponent $r$ of the two samples. As a consequence, we strongly recommend further investigations of this aspect for different sets of measurements and uncertainties and also for samples of stars different than the ones used here.

Moreover, as shown in Tables~(\ref{tab:correlation}), (\ref{tab:correlationsc}), and (\ref{tab:correlationlc}), the free parameters of the models are, in most of the cases, highly correlated. This may suggest to adopt different priors (e.\,g. non-uniform priors) for some of them in future analyses. The correlation is much more evident for RGs, possibly due to the larger range of the fundamental stellar properties in the LC sample considered.

Finally, as it is evident from many of the models investigated, the Sun is not necessarily a good average star from which to scale amplitudes for this sample of \textit{Kepler} stars, neither for the MS-dominated sample nor for the RGs. This outcome could be partially explained by the fact that the Sun is placed at the edge of our sample of measurements, as visible from Figure~\ref{fig:sample} (see also the discussion in Section~\ref{sec:model1}), and that its activity level is on average higher than that of the other stars \citep{Chaplin11act,Huber11}. Nevertheless, we stress that this result also relies on our choice of the set of uncertainties adopted for computing the likelihood. In fact, as discussed at the end of Section~\ref{sec:discuss_inference}, the uncertainties used are underestimated in most of the cases, even when considering errors on the variables of the models. This aspect may in fact produce a misleading solution to the inference problem and to the model selection. Since both the inference and the model comparison presented here rely on the assumption that these uncertainties are themselves error-free, in future work, it will be worth investigating how uncertainties affect the results by performing uncertainty-independent analyses that do not necessarily assume the adoption of a Gaussian likelihood function.

\section*{Acknowledgments}
EC acknowledges financial support from the PRIN-INAF 2010 {\it Asteroseismology: looking inside the stars with space- and ground-based observations.}
EC and JDR acknowledge the support of the FWO-Flanders under project O6260 - G.0728.11. 
Funding for the Stellar Astrophysics Centre is provided by The Danish National Research Foundation. The research is supported by the ASTERISK project (ASTERoseismic Investigations with SONG and Kepler) funded by the European Research Council (Grant agreement no.: 267864).

\bsp

\label{lastpage}


\begin{thebibliography}{99}
\bibitem[Andreon \& Hurn(2012)]{error1} Andreon, S., \& Hurn, M.~A.\ 2012, arXiv:1210.6232

\bibitem[Appourchaux et al.(1998)]{App98} Appourchaux, T., Gizon, L., \& Rabello-Soares, M.-C.\ 1998, A\&AS, 132, 107 

\bibitem[Appourchaux et al.(2012)]{App12} Appourchaux, T., Benomar, O., Gruberbauer, M., et al.\ 2012, A\&A, 537, A134 

\bibitem[Baglin et al.(2006)]{Baglin06} Baglin, A., Michel, E., 
Auvergne, M., \& COROT Team 2006, Proceedings of SOHO 18/GONG 2006/HELAS I, Beyond the spherical Sun, 624 

\bibitem[Ballot et al.(2011)]{Ballot11} Ballot, J., Barban, C., \& van't Veer-Menneret, C.\ 2011, A\&A, 531, A124 

\bibitem[Baudin et al.(2011)]{Baudin11} Baudin, F., Barban, C., Belkacem, K., et al.\ 2011, A\&A, 529, A84 

\bibitem[Bedding(2011)]{BeddingWS}
{Bedding}, T.~R. 2011, arXiv:1107.1723v1 [astro-ph.SR]

\bibitem[Bedding \& Kjeldsen(2003)]{BK03} Bedding, T.~R., \& Kjeldsen, H.\ 2003, PASA, 20, 203 

\bibitem[Bedding \& Kjeldsen(2008)]{BK08} Bedding, T.~R., \& Kjeldsen, H.\ 2008, 14th Cambridge Workshop on Cool Stars, Stellar Systems, and the Sun, ASPC, ed. {G.~van Belle}, 384, 21 

\bibitem[Bedding et al.(2007)]{Bedding07} Bedding, T.~R., 
Kjeldsen, H., Arentoft, T., et al.\ 2007, ApJ, 663, 1315 

\bibitem[Belkacem et al.(2011)]{Belkacem11a} Belkacem, K., Samadi, R., \& Goupil, M.~J.\ 2011, Journal of Physics Conference Series, 271, 012047 

\bibitem[Belkacem et al.(2012)]{Belkacem2012}
{Belkacem}, K., {Dupret}, M.~A., {Baudin}, F., {Appourchaux}, T., {Marques},
  J.~P., \& {Samadi}, R. 2012, A\&A, 540, L7

\bibitem[Belkacem(2012)]{Belkacem12b} Belkacem, K.\ 2012, SF2A-2012: Proceedings of the Annual meeting of the French Society of 
Astronomy and Astrophysics, 173  

\bibitem[Benomar et al.(2009)]{Benomar09} Benomar, O., Appourchaux, T., \& Baudin, F.\ 2009, A\&A, 506, 15

\bibitem[Bonanno et al.(2008)]{Bonanno08} Bonanno, A., Benatti, 
S., Claudi, R., et al.\ 2008, ApJ, 676, 1248 

\bibitem[Borucki et al.(2010)]{Borucki10} Borucki, W.~J., Koch, 
D., Basri, G., et al.\ 2010, Science, 327, 977 

\bibitem[Brown et al.(1991)]{Brown91} Brown, T.~M., Gilliland, 
R.~L., Noyes, R.~W., \& Ramsey, L.~W.\ 1991, ApJ, 368, 599 

\bibitem[Brown et al.(2011)]{Brown11} Brown, T.~M., Latham, 
D.~W., Everett, M.~E., \& Esquerdo, G.~A.\ 2011, AJ, 142, 112 

\bibitem[Campante et al.(2011)]{Campante11} Campante, T.~L., Handberg, R., Mathur, S., et al.\ 2011, A\&A, 534, A6 

\bibitem[Casagrande et 
al.(2010)]{Casagrande10} Casagrande, L., Ram{\'{\i}}rez, I., Mel{\'e}ndez, J., Bessell, M., \& Asplund, M.\ 2010, A\&A, 512, A54 

\bibitem[Chaplin et al.(2009)]{Chaplin09} Chaplin, W.~J., Houdek, G., Karoff, C., Elsworth, Y., \& New, R.\ 2009, A\&A, 500, L21 

\bibitem[Chaplin et al.(2011a)]{Chaplin11sc} Chaplin, W.~J., Kjeldsen, H., Christensen-Dalsgaard, J., et al.\ 2011a, Science, 332, 213 

\bibitem[Chaplin et al.(2011b)]{Chaplin11} Chaplin, W.~J., 
Kjeldsen, H., Bedding, T.~R., et al.\ 2011b, ApJ, 732, 54

\bibitem[Chaplin et al.(2011c)]{Chaplin11act} Chaplin, W.~J., 				
Bedding, T.~R., Bonanno, A., et al.\ 2011c, ApJ, 732, L5 

\bibitem[Christensen-Dalsgaard \& Frandsen(1983)]{CD83} Christensen-Dalsgaard, J., \& Frandsen, S.\ 1983, SoPh, 82, 469 

\bibitem[Corsaro et al.(2012a)]{Corsaro12a} Corsaro, E., Grundahl, F., Leccia, S., et al.\ 2012a, A\&A, 537, A9 

\bibitem[Corsaro et al.(2012b)]{Corsaro12b} Corsaro, E., Stello, D., Huber, D., et al.\ 2012b, ApJ, 757, 190 

\bibitem[D'Agostini(2005)]{error2} D'Agostini, G.\ 2005, arXiv:physics/0511182

\bibitem[Dziembowski \& Soszy{\'n}ski(2010)]{Dz10} Dziembowski, W.~A., \& Soszy{\'n}ski, I.\ 2010, A\&A, 524, A88 

\bibitem[{{Garc{\'{\i}}a} {et~al.}(2011){Garc{\'{\i}}a}, {Hekker}, {Stello},				
  {Guti{\'e}rrez-Soto}, {Handberg}, {Huber}, {Karoff}, {Uytterhoeven},
  {Appourchaux}, {Chaplin}, {Elsworth}, {Mathur}, {Ballot},
  {Christensen-Dalsgaard}, {Gilliland}, {Houdek}, {Jenkins}, {Kjeldsen},
  {McCauliff}, {Metcalfe}, {Middour}, {Molenda-Zakowicz}, {Monteiro}, {Smith},
  \& {Thompson}}]{Garcia11}
Garc{\'{\i}}a, R.~A., Hekker, S., Stello, D., et al.\ 2011, MNRAS, 414, L6 

\bibitem[Gilliland(2008)]{Gilliland08} Gilliland, R.~L.\ 2008, AJ, 136, 566			

\bibitem[Gilliland et al.(2010a)]{Gilliland10a} Gilliland, R.~L., 			
Jenkins, J.~M., Borucki, W.~J., et al.\ 2010a, ApJ, 713, L160 

\bibitem[Gilliland et al.(2010b)]{Gilliland10b} Gilliland, R.~L., 			
Brown, T.~M., Christensen-Dalsgaard, J., et al.\ 2010b, PASP, 122, 131  

\bibitem[Gruberbauer et al.(2012)]{Gruber12} Gruberbauer, M., 
Guenther, D.~B., \& Kallinger, T.\ 2012, ApJ, 749, 109 

\bibitem[Handberg \& Campante(2011)]{Handberg11} 
Handberg, R., \& Campante, T.~L.\ 2011, A\&A, 527, A56

\bibitem[Houdek et al.(1999)]{Houdek99} Houdek, G., Balmforth, N.~J., Christensen-Dalsgaard, J., \& Gough, D.~O.\ 1999, A\&A, 351, 582 

\bibitem[Houdek \& Gough(2002)]{Houdek02} Houdek, G., \& Gough, D.~O.\ 2002, MNRAS, 336, L65 

\bibitem[Houdek(2006)]{Houdek06} Houdek, G.\ 2006, Proceedings of SOHO 18/GONG 2006/HELAS I, Beyond the spherical Sun, 624  

\bibitem[Huber et al.(2009)]{Huber09} Huber, D., Stello, D., 
Bedding, T.~R., et al.\ 2009, Communications in Asteroseismology, 160, 74 

\bibitem[Huber et al.(2010)]{Huber10redgiant} Huber, D., Bedding, 
T.~R., Stello, D., et al.\ 2010, ApJ, 723, 1607

\bibitem[Huber et al.(2011a)]{Huber11a} Huber, D., Bedding, 
T.~R., Arentoft, T., et al.\ 2011a, ApJ, 731, 94 								

\bibitem[Huber et al.(2011b)]{Huber11} Huber, D., Bedding, 
T.~R., Stello, D., et al.\ 2011b, ApJ, 743, 143 								

\bibitem[Jenkins et al.(2010)]{Jenkins10} Jenkins, J.~M., 						
Caldwell, D.~A., Chandrasekaran, H., et al.\ 2010, ApJ, 713, L120 

\bibitem[Kass \& Wasserman(1996)]{Kass} Kass, R. E., \& Wasserman, L. 1996, J. Am. Stat. Association, 91, 1343

\bibitem[Kjeldsen \& Bedding(1995)]{KB95} Kjeldsen, H., \& Bedding, T.~R.\ 1995, A\&A, 293, 87

\bibitem[Kjeldsen \& Bedding(2011)]{KB11} Kjeldsen, H., \& Bedding, T.~R.\ 2011, A\&A, 529, L8 

\bibitem[Kjeldsen et al.(2005)]{K05} Kjeldsen, H., Bedding, 
T.~R., Butler, R.~P., et al.\ 2005, ApJ, 635, 1281 

\bibitem[Kjeldsen et al.(2008)]{K08} Kjeldsen, H., Bedding, 
T.~R., Arentoft, T., et al.\ 2008, ApJ, 682, 1370 

\bibitem[Koch et al.(2010)]{Koch10} Koch, D.~G., Borucki, W.~J., Basri, G., et al.\ 2010, ApJ, 713, L79 

\bibitem[Mathur et al.(2011)]{Mathur11} Mathur, S., Handberg, 
R., Campante, T.~L., et al.\ 2011, ApJ, 733, 95 

\bibitem[Michel et al.(2008)]{Michel08} Michel, E., Baglin, A., 			
Auvergne, M., et al.\ 2008, Science, 322, 558 

\bibitem[Michel et al.(2009)]{Michel09} Michel, E., Samadi, R., Baudin, F., et al.\ 2009, A\&A, 495, 979 		

\bibitem[Miglio et al.(2012)]{Miglio12} Miglio, A., Brogaard, 
K., Stello, D., et al.\ 2012, MNRAS, 419, 2077
 
\bibitem[Mosser et al.(2010)]{Mosser10} Mosser, B., Belkacem, K., Goupil, M.-J., et al.\ 2010, A\&A, 517, A22 

\bibitem[Mosser et al.(2012)]{Mosser12} Mosser, B., Elsworth, Y., Hekker, S., et al.\ 2012, A\&A, 537, A30	

\bibitem[Pinsonneault et al.(2012a)]{P12} Pinsonneault, M.~H., An, D., Molenda-{\.Z}akowicz, J., et al.\ 2012a, ApJS, 199, 30 

\bibitem[Pinsonneault et al.(2012b)]{Pcat12} Pinsonneault, M.~H., An, D., Molenda-Zakowicz, J., et al.\ 2012b, VizieR Online Data Catalog, 219, 90030 

\bibitem[Samadi et al.(2007)]{Samadi07} Samadi, R., Georgobiani, D., Trampedach, R., et al.\ 2007, A\&A, 463, 297  

\bibitem[Samadi et al.(2010)]{Samadi10} Samadi, R., Ludwig, H.-G., Belkacem, K., et al.\ 2010, A\&A, 509, A16 

\bibitem[Samadi et al.(2012)]{Samadi12} Samadi, R., Belkacem, K., Dupret, M.-A., et al.\ 2012, A\&A, 543, A120  

\bibitem[Schwarz(1978)]{schwarz} Schwarz, G., Ann. Statist. 6 461Ð464 (1978)

\bibitem[Stello et al.(2009a)]{Stello09a} Stello, D., Chaplin, 
W.~J., Basu, S., Elsworth, Y., \& Bedding, T.~R.\ 2009, MNRAS, 400, L80		

\bibitem[Stello et al.(2010)]{Stello10} Stello, D., Basu, S., Bruntt, H., et al.\ 2010, ApJL, 713, L182  		

\bibitem[Stello et al.(2011)]{Stello11} 							
Stello, D., Huber, D., Kallinger, T., et al.\ 2011, ApJ, 737, L10 

\bibitem[Still(2012)]{Still12} Still, M.~D.\ 2012, American 
Astronomical Society Meeting Abstracts, 220, \#419.01 

\bibitem[Trotta(2008)]{Trotta08} Trotta, R.\ 2008, Contemporary 
Physics, 49, 71

\bibitem[Verner et al.(2011)]{Verner11} Verner, G.~A., Elsworth, 
Y., Chaplin, W.~J., et al.\ 2011, MNRAS, 415, 3539 

\end{thebibliography}
\end{document}